\documentclass[preprint,pre,showpacs,floatfix,superscriptaddress]{revtex4}
\usepackage{graphicx}
\usepackage{dcolumn}
\usepackage{bm}
\usepackage{amsmath,amsfonts,amssymb,dsfont,color,bbm,tikz,bbold}
\usepackage{epstopdf}
\usepackage{dsfont}

\newcommand{\bra}[1]{\left\langle #1\right|}
\newcommand{\ket}[1]{\left| #1\right\rangle}
\newcommand{\ip}[2]{\left\langle #1 | #2\right\rangle}

\begin{document}

\title{Kicked-Harper model vs On-Resonance Double Kicked Rotor Model: From Spectral
Difference to Topological Equivalence}
\author{Hailong Wang}
\affiliation{Department of Physics and Center for Computational Science and Engineering,
National University of Singapore, 117542, Singapore}
\author {Derek Y.~H. Ho}
\affiliation{Department of Physics and Center for Computational Science and Engineering,
National University of Singapore, 117542, Singapore}
\author{Wayne Lawton}
\affiliation{ School
of Mathematics and Statistics, University of Western Australia, Perth, Australia}
\author{Jiao Wang} \email{phywangj@xmu.edu.cn}
\affiliation{Department of Physics and Institute of Theoretical Physics and Astrophysics,
Xiamen University, Xiamen 361005, China}
\author{Jiangbin Gong} \email{phygj@nus.edu.sg}
\affiliation{Department of Physics and Center for Computational Science and Engineering,
National University of Singapore, 117542, Singapore}
\affiliation{NUS Graduate School for Integrative Sciences and Engineering, Singapore
117597, Singapore}
\date{\today}

\begin{abstract}

Recent studies have established that, in addition to the well-known kicked Harper
model (KHM), an on-resonance double kicked rotor model (ORDKR) also has
Hofstadter's butterfly Floquet spectrum, with strong resemblance to the standard
Hofstadter's spectrum that is a paradigm in studies of the integer quantum Hall effect.
Earlier it was shown that the quasi-energy spectra of these two dynamical models
(i) can exactly overlap with each other if an effective Planck constant takes
irrational multiples of $2\pi$ and (ii) will be different if the same parameter takes
rational multiples of $2\pi$. This work makes some detailed comparisons between
these two models, with an effective Planck constant given by $2\pi M/N$, where
$M$ and $N$ are coprime and odd integers. It is found that the
ORDKR spectrum (with two periodic kicking sequences having the same kick strength)
has one flat band and $N-1$ non-flat bands whose largest width decays
in power law as $\sim K^{N+2}$, where $K$ is a kicking strength parameter. The
existence of a flat band is strictly proven and the power law scaling, numerically
checked for a number of cases, is also analytically proven for a three-band case.
By contrast, the KHM does not have any flat band and its band width scales
linearly with $K$. This is shown to result in dramatic differences in dynamical
behavior, such as transient (but extremely long) dynamical localization in ORDKR, which is absent in KHM.
Finally, we show that despite these differences, there exist simple extensions of
KHM and ORDKR (upon introducing
an additional periodic phase parameter) such that the resulting
extended KHM and ORDKR are actually topologically equivalent, i.e., they
yield exactly the same
Floquet-band Chern numbers and display topological phase transitions at the same
kick strengths. A theoretical derivation of this topological equivalence is provided.
These results are also of interest to our current understanding of quantum-classical
correspondence considering that  KHM and ORDKR
have exactly the same classical limit after a simple canonical transformation.

\end{abstract}

\maketitle

\section{Introduction}

As one important paradigm in the studies of quantum chaos and quantum-classical
correspondence, the kicked rotor (KR) model~\cite{KR} has received tremendous
theoretical and experimental interest in the last three decades~\cite{KR, Izrailev}.
For some experimental activities on KR within the last three years, we would like
to mention those listed in Ref.~\cite{recentKR}. A one-dimensional KR is described
by the Hamiltonian
\begin{eqnarray}
H_{\text{KR}}=p^{2}/2+K\cos(q)\sum_{n}\delta(t-nT)
\end{eqnarray}
in terms of dimensionless variables, where $p$ and $q$ are conjugate (angular)
momentum and angle variables, $K$ and $T$ are the kick strength and the period
of the $\delta$-kicks. The dynamical evolution of the system for a period from time
$nT+0^-$ to $(n+1)T+0^-$ can be expressed as a quantum map, which is given by
the following unitary Floquet operator
\begin{eqnarray}
U_{\text{KR}}=e^{-iT\frac{p^2}{2\hbar}}e^{-i\frac{K}{\hbar}\cos(q)}.
\label{KR}
\end{eqnarray}
For our considerations below, we confine ourselves to a rotor Hilbert space defined by
 the periodic boundary condition in $q$, with $q\in[0,2\pi)$. The
Hilbert space can then be represented by the eigenfunctions $\{|m\rangle\}$ of
$p$, with $p|m\rangle=m\hbar|m\rangle$, $\langle q|m\rangle=\exp(imq)/\sqrt{2\pi}$,
$m$ being an integer, and $\hbar$ being a dimensionless effective Planck constant.
Through extensive numerical simulations and mathematical analysis, it is now well
known that in general the KR dynamics can be classified into two
categories~\cite{Izrailev}. For an irrational (hence generic) value of  $T\hbar/(2\pi)$
the system can diffuse in (angular) momentum space only for a short time due to ``dynamical localization'', regardless
of the kick strength.  This
hints at a discrete spectrum of $U_{\text{KR}}$ and is closely related to Anderson
localization~\cite{ATr}. On the other hand, for $T\hbar/(2\pi)$ being a rational multiple of
$2\pi$ (except for odd multiples of $2\pi$), $U_{\text{KR}}$ has continuous bands:
A time-evolving state would keep spreading out in (angular) momentum space ballistically.
This category of dynamics was termed as ``quantum resonance''~\cite{QR}.


Another important quantum chaos model is the kicked Harper model
(KHM)~\cite{leboeuf, KH1, KH2}, originally introduced in Ref.~\cite{zas} as an approximation
  of the problem of kicked charges in a magnetic field.  Remarkably, the KHM and even a whole class of its generalized
  versions were shown to be equivalent to the problem of a charge kicked periodically
  in the presence of a magnetic field \cite{dana-pla}.
   The associated KHM quantum map for each period is
given by
\begin{eqnarray}
U_{\text{KHM}}=e^{-i\frac{L}{\hbar}\cos(p)}e^{-i\frac{K}{\hbar}\cos(q)},
\label{KHM}
\end{eqnarray}
with $L$ being an additional system parameter. Throughout we assume the KHM
is also treated in the same Hilbert space as the KR and is quantized on a
rotor Hilbert space. The dynamics of KHM differs from that of KR as described above
in several aspects. For
example, for all irrational values of $\hbar/(2\pi)$, the system
in general tends to delocalize (localize) in (angular) momentum space for $K>L$
($K<L$)~\cite{KH2}. Of particular interest is the symmetric case of $K=L$, for
which the quasi-energy spectrum of $U_{\text{KHM}}$ is fractal-like in general.
Scanning the spectrum collectively  for fixed $K/\hbar=L/\hbar$ versus a varying
$\hbar$ forms a pattern that resembles the Hofstadter's butterfly
spectrum~\cite{hofstadter}, a paradigm in studies of the integer quantum Hall
effect. The associated dynamics is extended in general and may be connected
with the fractal dimensions of the Floquet spectrum.

Given the above-mentioned differences between KR and KHM, the work of
Ref.~\cite{JJ08} by two of the authors emerged somewhat unexpectedly. There
it was shown that a variant of KR also has Hofstadter's butterfly spectrum. In
particular, motivated by the double-kicked rotor model studied both experimentally and theoretically
in Ref.~\cite{DKRM}, which is a special case of ``multiple KR's" first introduced in Ref.~\cite{dana-flat-band2},
Ref.~\cite{JJ08} studied  a double-kicked model under  a quantum-resonance condition. For a total period of
$\tau$ ($\tau>1$), a double kicked rotor model is associated with two periodic $\delta$-kicks
of strengths $K$ and $L$, separated by a time interval set to be unity, yielding the following Floquet
operator
\begin{eqnarray}
U_{\text{DKR}}=e^{-i(\tau-1)\frac{p^2}{2\hbar}}e^{-i\frac{K}{\hbar}\cos(q)}
e^{-i\frac{p^2}{2\hbar}}e^{-i\frac{L}{\hbar}\cos(q)}.
\label{DKR}
\end{eqnarray}
In Ref.~\cite{JJ08}, $\tau$ is chosen to satisfy the quantum resonance condition
$\tau\hbar=4\pi$. Then $e^{-i\tau\frac{p^2}{2\hbar}}=1$ due to the discreteness
of the momentum eigenvalues. This leads us to an on-resonance double kicked
rotor model (ORDKR), whose Floquet operator is given by \cite{JJ07}:
\begin{eqnarray}
U_{\text{ORDKR}}=e^{i\frac{p^2}{2\hbar}}e^{-i\frac{K}{\hbar}\cos(q)}
e^{-i\frac{p^2}{2\hbar}}e^{-i\frac{L}{\hbar}\cos(q)}.
\label{ORDKR}
\end{eqnarray}
Note that we have deliberately used symbols $K$ and $L$ in both $U_{\text{KHM}}$
and $U_{\text{ORDKR}}$ because in this paper, parameter $K$ or parameter $L$ from both models will always be assigned the same value. Experimental realization of such an ORDKR propagator in atom optics is possible
by loading a Bose-Einstein-Condensate (BEC) in a kicking optical lattice, with the initial quasi-momentum spread
of the BEC negligibly small as compared with the recoil momentum of the optical lattice \cite{exp-note}.
Interestingly, for $\hbar$ being an irrational multiple of $2\pi$, the ORDKR and the
KHM share the same quasi-energy spectrum~\cite{JMP,GuarDKR}.

Our main plan for this paper is to make some detailed comparisons between KHM
and ORDKR as two closely related dynamical models, both possessing Hofstadter's
butterfly spectrum.  Our motivations are as follows. First of all, in
Refs.~\cite{JJ08,JMP}, it was shown that $U_{\text{ORDKR}}$ and $U_{\text{KHM}}$
have different spectra
if $\hbar$  is a rational multiple of $2\pi$. On the other hand,
as $\hbar/(2\pi)$ approaches an arbitrary irrational number, the spectral difference between $U_{\text{ORDKR}}$ and $U_{\text{KHM}}$,
which is characterized by a Hausdorff metric in Ref.~\cite{JMP}, was shown to approach
zero.
It is therefore highly worthwhile looking into the
actual spectral differences for rational values of $\hbar/(2\pi)$, because, up to a classical canonical transformation, ORDKR
and KHM have exactly the same classical limit~\cite{JMO} (obtained by letting
$\hbar$ approach zero while fixing $K/\hbar$ and $L/\hbar$). Indeed, given their
equivalence in the classical limit, the spectral differences we analyze constitute
beautiful examples to illustrate how quantization of classically equivalent systems
may lead to remarkable system-specific consequences. Second, by working on the
details we hope to find some clues as to why the dynamics of ORDKR can be so
different from that of KHM.  We indeed succeed in doing this, finding that even
on a qualitative level, the Floquet bands of ORDKR behave much differently from
that of KHM, for $\hbar=2\pi M/N$, with $M$ and $N$ being coprime and both odd.
In particular, we shall prove the existence of a flat Floquet band \cite{dana-flat-band1,dana-flat-band2}
for ORDKR with $K=L$, which
may be of interest to current studies of strongly correlated condensed-matter systems
with an almost flat energy band \cite{flat-band}. The existence of a flat Floquet band has been shown elsewhere to be important in explaining the intriguing exponential quantum spreading dynamics in ORDKR~\cite{JiaoPRL11,Hailongwork}.
Third, motivated by recent interests in topological characterization of periodically
driven  systems~\cite{derek12,floquetTI} and given the interesting relationship of the two models described previously,
we ask whether, after all, ORDKR and
KHM have any interesting topological connections. Based on our numerical and analytical
studies, the answer is yes and we shall claim that ORDKR
and KHM are topologically equivalent in the sense that their extended Floquet bands (obtained upon introducing a phase shift parameter defined in Sec.~III) always have the same band Chern numbers.

This paper is organized in the following order. In Sec.~II we present detailed results
regarding a spectral comparison between KHM and ORDKR, for $K=L$, and
$\hbar=2\pi M/N$ with $M$ and $N$ being coprime and odd integers. Numerical
findings will be described first, followed by analytical considerations when possible
(e.g., band width scaling for a three-band case and the general proof of a flat band
for ORDKR).  The implications of peculiar spectral properties of ORDKR for its
dynamics are also discussed via some numerical studies. In Sec.~III we study the
KHM and ORDKR by extending them to accommodate a new periodic parameter
and demonstrating the topological equivalence of the resulting extended models. Section IV concludes this paper.


\section{Spectral differences and their dynamical implications}

\subsection{Summary of main numerical findings}

As far as numerics are concerned, the spectrum of the unitary operators can be
obtained in a straightforward manner. For completeness we describe some details
here. The key step is to take advantage of the periodic property of
$U_{\text{KHM}}$ or $U_{\text{ORDKR}}$ in the (angular) momentum space, which arises
naturally for $\hbar$ being a rational multiple of $2\pi$. Denote $U$ to refer to
either $U_{\text{KHM}}$ or $U_{\text{ORDKR}}$. Letting
$U_{j,k}\equiv\bra{j}U\ket{k}$, one easily finds $U_{j+N,k+N}=U_{j,k}$ for
$\hbar=2\pi M/N$. This indicates a unit cell in (angular) momentum space, with a size of $N$.
The spectrum is then equivalent to that of a reduced $N\times N$ matrix
$\tilde{U} (\varphi)$, whose elements are given by
$[\tilde{U}(\varphi)]_{j,k}=\sum\limits_{l} e^{il\varphi}U_{j,k+lN}$, with $\varphi\in [0,2\pi)$
being the Bloch phase in momentum space and $l$ running over all integers. As
off-diagonal elements of $U_{j,k}$ decay exponentially, the summation in
$\sum\limits_l e^{il\varphi}U_{j,k+lN}$ can be truncated safely at certain large enough
value of $|l|$ (in our analytical studies below, we do not do such truncations).
Numerical results are then checked by further increasing the truncation radius.
Once $\tilde{U}(\varphi)$ is numerically obtained, the standard diagonalization
algorithm for a unitary matrix can be exploited to obtain $N$ values of quasi-energy
$\epsilon$. By varying $\varphi$ in $[0,2\pi)$ we have $N$ Floquet bands.

\begin{figure}
\centering
\rotatebox{90}{
\includegraphics[trim=0cm 11cm 20cm 0cm,clip=true,height=!,width=8cm]{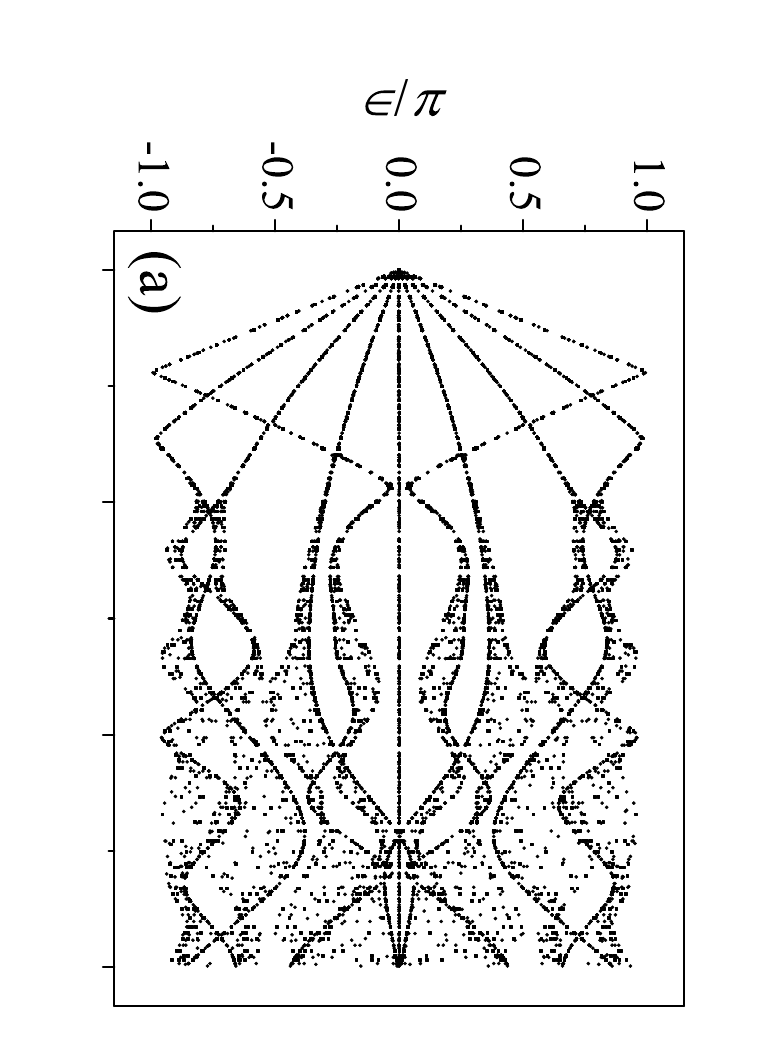}}
\vskip-1.0cm
\rotatebox{90}{
\includegraphics[trim=0cm 11cm 20cm 0cm,clip=true,height=!,width=8cm]{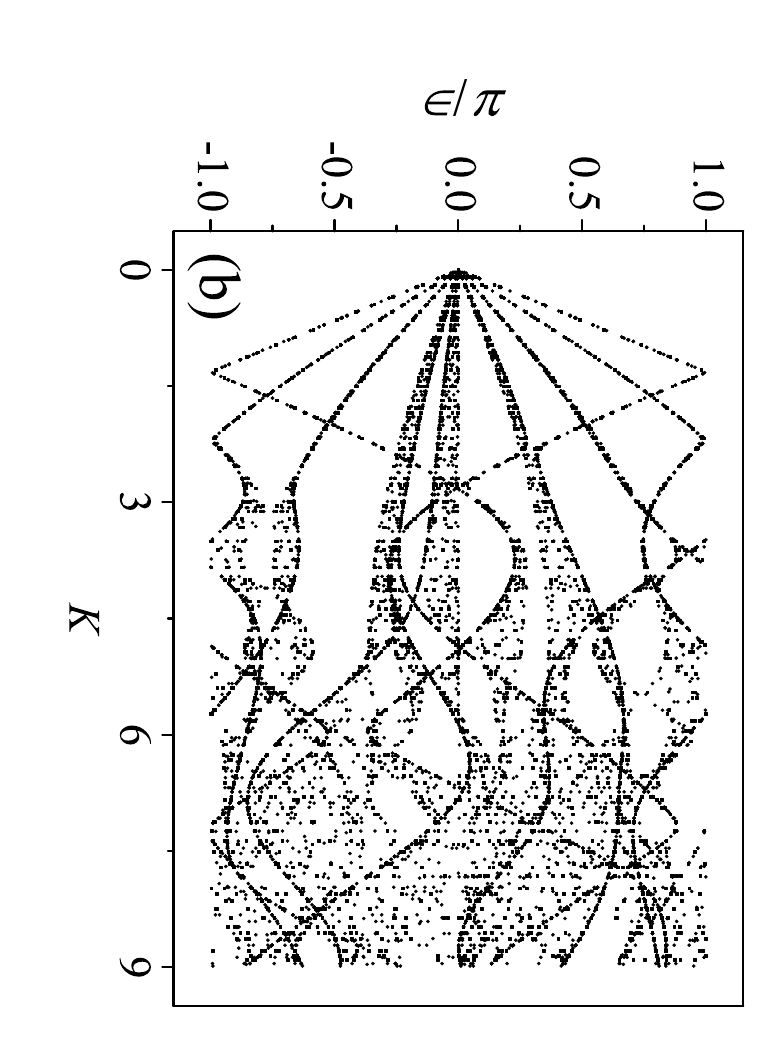}}
\vskip-0.5cm
\caption{The quasi-energy bands versus the kick strength $K=L$ for an effective
Planck constant $\hbar=2\pi M/N$, with $M=1$ and $N=9$, for ORDKR in panel
(a) for KHM in panel (b). Note that for ORDKR, there is a straight line lying in the
middle of the spectrum, indicating the existence of a flat band.  Here and in all other figures, all
plotted quantities are in dimensionless units.}
\label{squid}
\end{figure}

In Fig.~\ref{squid} we show our obtained quasi-energy values of $U_{\text{ORDKR}}$ and $U_{\text{KHM}}$
as a function of the kick strength $K$. Though for each fixed value of $K$,
we only show the quasi-energy values for a limited number of Bloch phase choices,
the locations of the bands, the band width, and a few avoided band crossings can
already be seen clearly for not too large values of $K=L$. In particular,  at $N=9$,
nice Floquet bands can be identified clearly for both ORDKR and KHM, though for
very large values of $K$ the merging of the bands does occur.

Spectral differences between $U_{\text{ORDKR}}$ and $U_{\text{KHM}}$ in the shown example are also
obvious. Based on the results shown in Fig.~\ref{squid}, we have carried out
extensive numerical investigations for other cases with $\hbar=2\pi M/N$, with
$M$ and $N$ coprime and both odd.   Some of the main features are presented and commented on below.

First, the band structure of $U_{\text{ORDKR}}$ is symmetric with respect to the zero
quasi-energy axis, which is however not the case for $U_{\text{KHM}}$. This interesting
symmetry is absent in both $U_{\text{KHM}}$ and $U_{\text{KR}}$.  We shall prove this property below.

Second, consistent with the above-mentioned symmetry, $U_{\text{ORDKR}}$ is seen to have a
flat band with $\epsilon=0$. By flat band, we mean that this quasi-energy value is
independent of the Bloch phase $\varphi$.  So the overall picture is that the $N$
bands can be classified into $(N-1)/2$ pairs, with each pair having opposite
quasi-energy values, plus a flat band in the middle. Again, this is not the case for
$U_{\text{KHM}}$.  The existence of a flat Floquet band was previously observed
in studies of the quantum antiresonance phenomenon in kicked systems \cite{dana-flat-band1,dana-flat-band2}.
However, unless in the case of $N=1$ ($M$ odd) that also corresponds to a quantum antiresonance condition,
here the flat band of $U_{\text{ORDKR}}$ coexists with other nonflat bands. This coexistence of a flat band with nonflat bands
constitutes an interesting feature.
As a side note, Ref.~\cite{GuarORKR} suggested that for a KR defined in this paper
under the quantum
resonance condition of any order (i.e., $T\hbar=4\pi M/N$, with $M$ and $N$ arbitrary coprime integers), none of the Floquet
bands of $U_{\text{KR}}$ is flat. So the existence of one single flat band of $U_{\text{ORDKR}}$ is also
remarkable as compared with $U_{\text{KR}}$.


\begin{figure}
\centering
\includegraphics[trim=0cm 0cm 11cm 20.5cm,clip=true,height=! ,width=9cm]{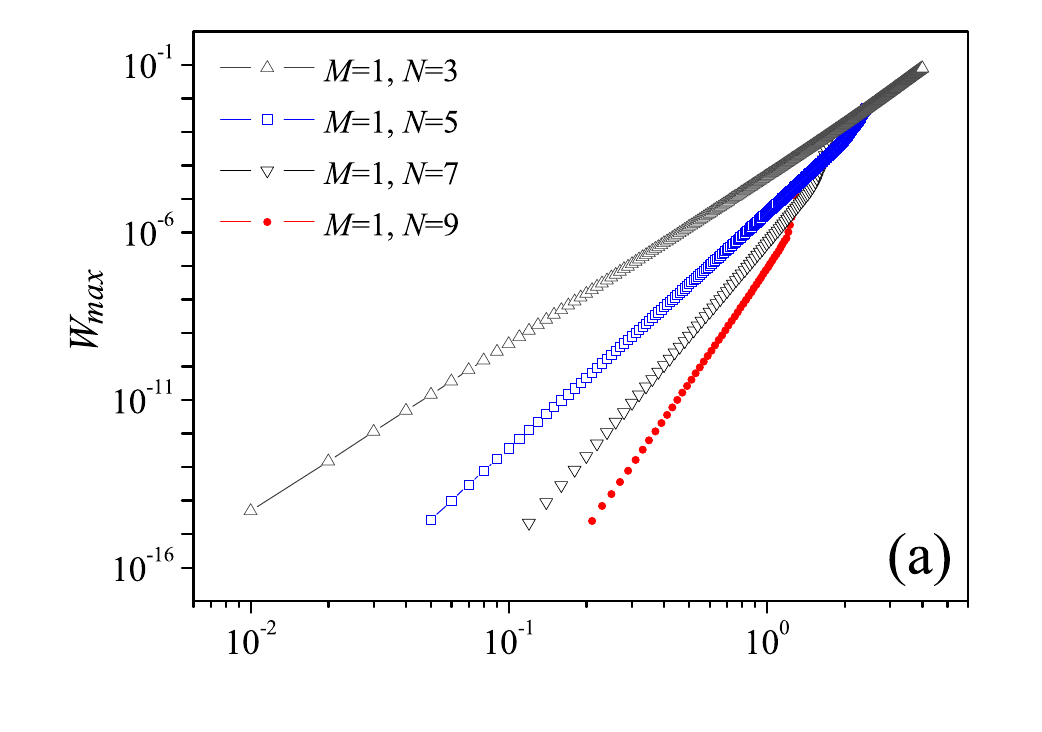}
\vskip-0.90cm
\includegraphics[trim=0cm 0cm 11cm 20cm,clip=true,height=! ,width=9cm]{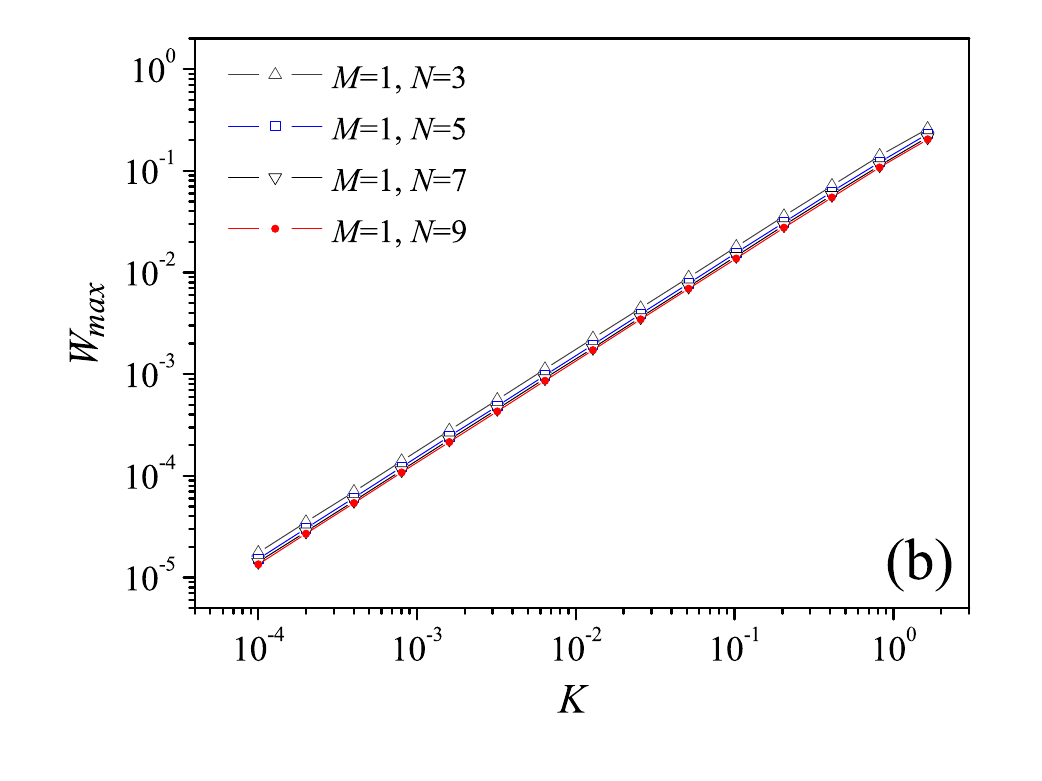}
\vskip-0.5cm
\caption{(color online) The bandwidth of the widest band, denoted $W_{max}$,
as a function of the kick strength parameter $K=L$ for (a) ORDKR and (b) KHM.
In both panels, the effective Planck constant $\hbar=2\pi M/N$ with $M=1$ and
$N=3, 5, 7$, and $9$, respectively. In the former case $W_{max}\sim K^{N+2}$
as $K\to 0$ but in the latter case it always scales linearly with $K$.}
\label{BW}
\end{figure}

Third, the largest bandwidth of the other $N-1$ non-flat Floquet bands of ORDKR
scales with $K$ as $\sim K^{N+2}$, in the limit of $K\to 0$. In sharp contrast, the
bandwidths of KHM scale with $K$ linearly. Representative numerical results are
shown in Fig.~\ref{BW}, where the bandwidth of the widest band is plotted against
small values of $K$, for $\hbar=2\pi M/N$, with $M=1$, $N=3,5,7,9$.  The power
law decay of the ORDKR bandwidth in the form of $\sim K^{N+2}$ can be clearly
identified, whereas the bandwidth of KHM remains a linear function of $K$,
irrespective of the value of $N$.
This being the case, in the small $K$ regime ($K<<1$), the maximum bandwidths of ORDKR is
$K^{N+1}$ times narrower than that of KHM.


\subsection{Flat band and Band symmetry in ORDKR}

Flat bands in solid-state systems are of vast interest in condensed matter physics
because they offer new opportunities for understanding strongly correlated
systems without Landau levels.  For this reason the existence of a flat band
in a periodically driven system can be useful, too. To further understand the
flat band of ORDKR,  we present a theoretical proof in this subsection. In doing
so we shall also prove the band symmetry noted above. We shall also discuss
how an eigenstate on a flat band, which is infinitely degenerate, may be
numerically found.

For $\hbar=2\pi M/N$ with $M$ and $N$ being coprime integers, the spectrum
becomes that of a reduced $N\times N$ Floquet matrix with elements
$[{\tilde U}_{\text{ORDKR}}(\varphi)]_{n,m}=\sum_{l=-\infty}^{\infty}
\bra{n}{\hat U}_{\text{ORDKR}}\ket{m+l\times N} e^{il\varphi}$.
After performing some necessary integrals and using the fact that both $M$ and
$N$ are odd, one can express $\left[\tilde{U}_{\text{ORDKR}}(\varphi)\right]_{n, m}$ as a
summation of finite terms (see Appendix for details).  In the following discussions
regarding the existence of a flat band and the band inversion symmetry,  we shall restrict ourselves to the cases of $K=L$
(note however, in the next section,  the notation introduced here will be extended to the cases with $K\ne L$).
We first introduce diagonal unitary matrices
$D_{\varphi}$, $D_{1}$, $D_{K}$ and unitary matrix $F$, with matrix elements
$(D_{\varphi})_{n,m}=e^{-in\frac{\varphi}{N}}\delta _{n,m}$,
$(D_{1})_{n,m}=e^{i\frac{2\pi-\hbar}{2}n^2}\delta _{n,m}$,
$(D_{K})_{n,m}=e^{-i\frac{K}{\hbar}\cos(\frac{2\pi}{N}n-\frac{\varphi}{N})}\delta _{n,m}$
and $F_{m,n}=\frac{1}{\sqrt N}e^{i\frac{2\pi}{N}mn}$, where indices $m$ and $n$
take values $0,1,\cdots,N-1$. Note that in obtaining our expression for $D_1$, we
made use of the fact that $e^{i n^{2} \pi} = e^{i n \pi}$.
 We then have the following compact form for the reduced Floquet matrix
\begin{equation}
{\tilde U}_{\text{ORDKR}}(\varphi)=D_{\varphi}^{\dagger} D_{1}^{\dagger} F^{\dagger} D_{K}^{\dagger}FD_{1}F^{\dagger}D_{K}FD_{\varphi}.
\end{equation}

To prove that there is a flat band for ORDKR, we show that
$\tilde{U}_{\text{ORDKR}}(\varphi)$ has an eigenvalue equal to one, regardless of
the value of $\varphi$. Consider then a matrix ${\tilde U}'_{\text{ORDKR}}(\varphi)$
transformed from $\tilde{U}_{\text{ORDKR}}(\varphi)$ by a unitary operation
$FD_{\varphi}$, which takes the form
\begin{equation}
{\tilde U}'_{\text{ORDKR}}(\varphi)=\left(F D_1^\dagger F^\dagger\right)D_{K}^\dagger\left(FD_1F^\dagger\right)D_{K}.
\end{equation}
The eigenvalue equation of ${\tilde U}'_{\text{ORDKR}}(\varphi)$ may be rewritten as
\begin{equation}
(BD_{K}-\lambda D_{K}B)|x\rangle=0,
\label{neweigeneq}
\end{equation}
where $B\equiv FD_1F^{\dag}$, $|x\rangle$ denotes an eigenvector, and $\lambda$
is an eigenvalue of ${\tilde U}_{\text{ORDKR}}(\varphi)$. Detailed calculations show
that $B$ is a symmetric matrix (see Appendix for details) and since $D_K$ is a
diagonal matrix, $(B D_K-D_K B)$ must be an antisymmetric matrix of odd dimension.
It immediately follows $\text{Det}(B D_K-D_K B)=0$. Thus, regardless of the Bloch
phase $\varphi$, $\lambda=1$ is a permissible solution to Eq.~(\ref{neweigeneq}).
We have thus shown that ${\tilde U}_{\text{ORDKR}}(\varphi)$ always has a unity
eigenvalue or zero quasi-energy for $\hbar=2\pi M/N$.  This is nothing but the
existence of a flat Floquet band.

Our considerations above also lead us to a proof of the band inversion symmetry of
ORDKR for odd $M$ and $N$. Specifically, because
$(BD_K-\lambda D_KB)^{\text{T}}=(D_KB-\lambda BD_K)=-\lambda(BD_K-\lambda^{-1}D_KB)$,
we see that if $\text{Det}(BD_K-\lambda D_KB)=0$, then
$\text{Det}(BD_K-\lambda^{-1}D_KB)=0$ as well.  That is, both $\lambda$ and
$\lambda^{-1}$ are solutions to the eigenvalue equation of Eq.~(\ref{neweigeneq}).
As such, if we have a quasi-energy $\epsilon=i\ln\lambda$, we must have
$i\ln\lambda^{-1}=-\epsilon$ in the spectrum. This completes our proof of the
inversion symmetry of the ORDKR.

\begin{figure}
\centering
\rotatebox{90}{
\includegraphics[trim=0cm 11cm 18.5cm 0cm,clip=true,height=! ,width=8cm]{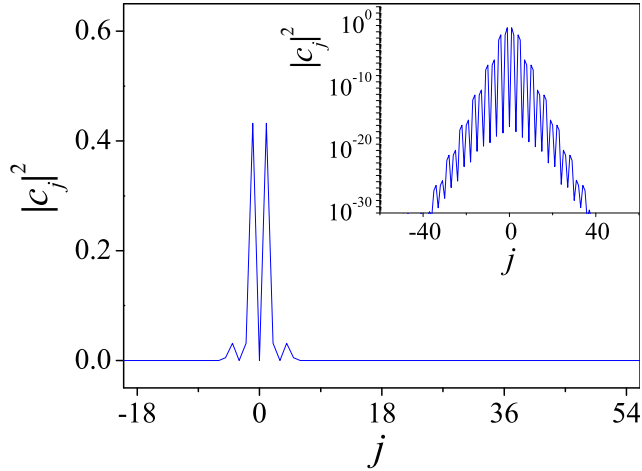}}
\vskip-0.5cm
\caption{(color online) A localized eigenstate $|\psi\rangle=\sum_j c_j |j\rangle$
associated with the flat-band in the on-resonance double kicked rotor model for
$K=3$ and $\hbar=2\pi /3$. The insert is the same but in semi-log scale.}
\label{eigenstate}
\end{figure}

A flat band is infinitely degenerate as states on the band can still have a continuous
Bloch phase $\varphi$.  Due to such an independence upon the Bloch phase,
the band dispersion relation directly yields a zero group velocity in the (angular) momentum space, thus indicating
a zero mobility in the (angular) momentum space.   Further,
the infinite degeneracy allows us to construct a flat-band eigenstate
that is localized in the (angular) momentum space (though the Floquet operator itself is
periodic in momentum with a period $N\hbar$). It is interesting to outline a simple
approach to the construction of flat-band states. It is found that, highly localized
flat-band states can be obtained by directly truncating the full Floquet matrix
${U}_{\text{ORDKR}}(\varphi)$ to a small size, such that there is only one eigenstate
whose eigenvalue is real and still equals to unity (thus not affected by the truncation).
Other localized states on the flat band can be obtained by shifting it by a multiple
of N sites, or by superimposing these states localized at different locations.
Figure~\ref{eigenstate} depicts one computational example of a flat-band eigenstate
strongly localized in the (angular) momentum space. We have checked that if we use a flat-band
state we constructed as the initial state for time evolution, then indeed this state
does not evolve with iterations of our ORDKR quantum map. This situation is more subtle than
the quantum antiresonance phenomenon \cite{dana-flat-band1, dana-flat-band2}:
for ORDKR with multiple bands, only special states prepared on the single flat band can remain localized,
whereas in the case of quantum antiresonance an arbitrary state should remain localized.

\subsection{A theoretical bandwidth result and its dynamical consequence}

For $\hbar=2\pi M/N$ with $M$ and $N$ being coprime integers, the reduced
$N\times N$ Floquet matrices $\tilde{U}_{\text{ORDKR}}(\varphi)$ and
$\tilde{U}_{\text{KHM}}(\varphi)$ (see our general expressions in the Appendix)
can be obtained analytically. To further understand and confirm the bandwidth
scaling of ORDKR and KHM, we have also carried out analytical studies for a
three-band case, with $K=L$ and $\hbar=2\pi/3$.

For ORDKR, the three eigenvalues are found to be $1$ and $e^{\pm i\epsilon(\varphi)}$,
where $\epsilon(\varphi)\equiv\arccos[\frac{1}{2}\text{Tr}\tilde{U}(\varphi)-\frac{1}{2}]$.
One finally finds
$\epsilon(\varphi)=\arccos\{\frac{1}{3}[2\cos(\frac{\sqrt{3}K}{2\hbar}\sin\frac{\varphi}{3})
\cos(\frac{3K}{2\hbar}\cos\frac{\varphi}{3})+\cos(\frac{\sqrt{3}K}{\hbar}\sin\frac{\varphi}{3})]\}$
where $\hbar=\frac{2\pi}{3}$. For $K<1$ it can be shown that the edges of the
band correspond to $\varphi=0(\pi)$ and $\varphi=\frac{\pi}{2}(\frac{3\pi}{2})$.
The bandwidth can thus be determined to be
$\arccos\left\{\left[\cos(\frac{\sqrt{3}K}{\hbar})+2\cos(\frac{\sqrt{3}K}{2\hbar})\right]/3\right\}
-\arccos\left[\frac{1}{3}+\frac{2}{3}\cos\left(\frac{3K}{2\hbar}\right)\right]$.
Taylor expanding this expression for the bandwidth, we find the first nonzero term
to be $\frac{\sqrt{6}}{1280}(\frac{K}{\hbar})^{5}$, a clear power-law scaling of $K^{5}$.

For KHM, the eigenvalues can be deduced from the equation
$\text{Det}\left({\tilde U}_{\text{KHM}}(\varphi)-\lambda\right)=0$.
The resulting explicit expression of the eigenvalue equation is
\begin{equation}
{\lambda}^3-3re^{i\theta}{\lambda}^2+3re^{-i\theta}\lambda-1=0,
\label{eigvalequ1}
\end{equation}
where
$re^{i\theta}=\frac{1}{9}\left(e^{-i\frac{K}{\hbar}}+2e^{i\frac{K}{2\hbar}}\right)
\left(e^{-i\frac{K}{\hbar}\cos{\frac{\varphi}{3}}}+2e^{i\frac{K}{2\hbar}
\cos{\frac{\varphi}{3}}}\cos(\frac{\sqrt{3}K}{2\hbar}\sin{\frac{\varphi}{3}})\right)$.
Note that all eigenvalues are in the form of $\lambda\equiv e^{-i\epsilon}$,
since the reduced Floquet matrix is always unitary. The three eigenvalues are
found to be
$e^{-i\epsilon_1}=re^{i\theta}+(re^{2i\theta}-e^{-i\theta})\frac{r}{z}+z$,
$e^{-i\epsilon_2}=re^{i\theta}+e^{-\frac{2i\pi}{3}}(re^{2i\theta}-
e^{-i\theta})\frac{r}{z}+e^{\frac{2i\pi}{3}}z$ and
$e^{-i\epsilon_3}=re^{i\theta}+e^{\frac{2i\pi}{3}}(re^{2i\theta}-
e^{-i\theta})\frac{r}{z}+e^{-\frac{2i\pi}{3}}z$ where
$z=\left(\frac{1}{2}-\frac{3}{2}r^2+r^3e^{3i\theta}+\sqrt{\frac{1}{4}-
\frac{3}{2}r^2+2r^3\cos(3\theta)-\frac{3}{4}r^4}\right)^{\frac{1}{3}}$.
For $K<1$, the edges of the band correspond to $\varphi=0$ and $\varphi=\pi$.
The band width can thus be determined to be
$W_1=|\epsilon_1(\varphi=0)-\epsilon_1(\varphi=\pi)|$,
$W_2=|\epsilon_2(\varphi=0)-\epsilon_2(\varphi=\pi)|$ and
$W_3=|\epsilon_3(\varphi=0)-\epsilon_3(\varphi=\pi)|$.
Taylor expanding the expressions of eigevalues for $K\ll 1$ and keeping the
lowest order in $K$, we have
$W_1\approx\sqrt{2}\sin(\frac{\pi}{12})\frac{K}{\hbar}$,
$W_2\approx\left(\sqrt{\frac{3}{2}}-1\right)\frac{K}{\hbar}$ and
$W_3\approx\left(\sqrt{2}\cos(\frac{\pi}{12})-\sqrt{\frac{3}{2}}\right)\frac{K}{\hbar}$,
a clear linear scaling of K.

\begin{figure}

\centering
\includegraphics[trim=0cm 0cm 11cm 13.5cm,clip=true,height=!,width=11cm]{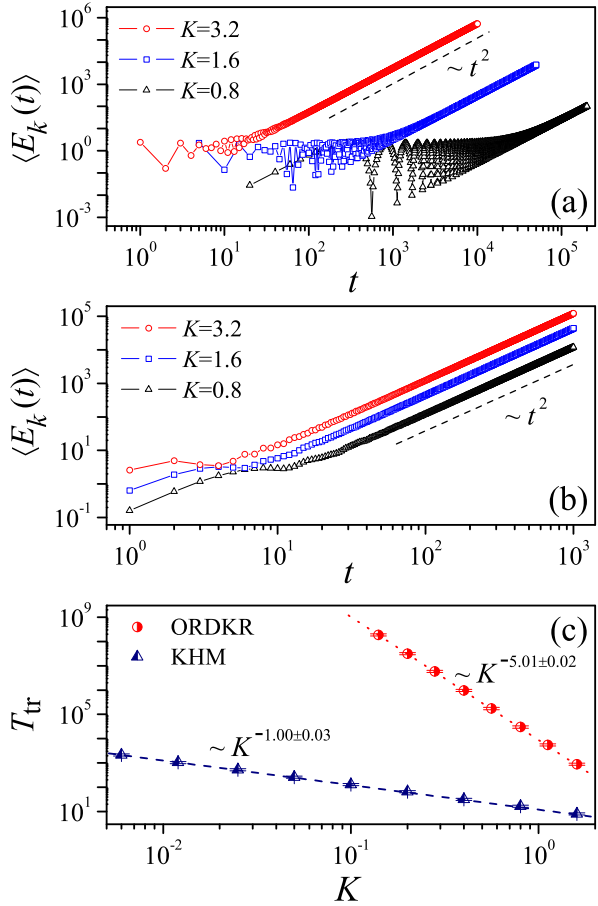}
\vskip-0.0cm
\caption{(color online) Panels (a) and (b) depict the expectation value of system's kinetic energy versus time $t$
(measured as the number of quantum maps iterated), with $\hbar=2\pi/3$ and
the initial state given by $|0\rangle$, for three values of kick strength $K=L$, with
(a) for ORDKR and (b) for KHM. For a small value of $K$, the
kinetic energy of ORDKR or KHM is seen to be localized for a long while before it starts
to increase ballistically.  Panel (c) shows how the time scale of this initial transient stage, denoted
$T_{\text{tr}}$,
scales with $K$: the scaling is found to be $\sim K^{-5}$ for ORDKR but $\sim K^{-1}$ for KHM,
which is consistent with our analysis of the respective bandwidth
power-law scaling with $K$.}
\label{p2t}
\end{figure}

The very fast decay of the Floquet bandwidth of ORDKR suggests that in a
considerable range of $K$ the bandwidths will be very narrow. In other words,
for a small $K$, all the Floquet bandwidths would be effectively zero for a
reasonably long time scale. Therefore, when it comes to the dynamical evolution
of the system, effectively the system will not feel its continuous Floquet
spectrum and hence displays localization behavior, for a time scale inversely
proportional to the bandwidths.  We call this the time scale of transient
dynamical localization and denote it by $T_{\text{tr}}$. We then have
$T_{\text{tr}}\sim K^{-(N+2)}$.  The overall expectation is the following: within
$T_{\text{tr}}$, ORDKR displays localization in the (angular) momentum space, but
afterwards it begins to show ballistic behavior in the (angular) momentum space.
Because of the power law scaling, the intriguing time scale $T_{\text{tr}}$
can be very sensitive to a change in the kick strength $K$. Our numerical
calculations indeed confirm this.  Figure~\ref{p2t}(a) shows an example of
the dynamics of the kinetic energy of ORDKR, starting from an initial state
with zero momentum.  In all three of the shown cases, the kinetic energy
is seen to freeze over a time scale before it starts to increase ballistically.
The time scale of the freezing stage is shown to increase rapidly as we
decrease the value of $K$.  As a comparison, Figure~\ref{p2t}(b)
shows the parallel dynamics of KHM, for the same three values of $K$.
There it is seen that the transient stage of localization is only weakly
dependent upon $K$, which is again consistent with the linear
$K$-dependence of the bandwidth of KHM.
Quantitatively, the
transient localization time scale $T_{\text{tr}}$ is numerically determined
from the duration of kinetic energy freezing.  The $T_{\text{tr}}$ thus
obtained numerically and shown in Fig.~\ref{p2t}(c)
indeed satisfies the scaling $T_{\text{tr}}\sim K^{-(N+2)}$ for ORDKR, which is in sharp contrast
to the $T_{\text{tr}}\sim K^{-1}$ scaling for KHM.  The results here can also be understood as a
quantitative explanation of our earlier finding of transient dynamical
localization in Ref.~\cite{JJ07}.  For future experiments, the observation
of the aforementioned scaling of $T_{\text{tr}}$ versus $K$ may serve as the first piece of evidence
of a successful realization of an ORDKR.

\section{Topological Equivalence between ORDKR and KHM}

In this section, we devote ourselves to a detailed comparison of the Floquet
band topologies of ORDKR and KHM. We first describe our motivation and
introduce new notation. Next, we report numerical findings of the Floquet
band topological numbers of both models. Finally, an exact analytical proof of
the topological equivalence between ORDKR and KHM is presented.

\subsection{Motivation and Notation}

One early study \cite{leboeuf} suggested that topological properties of the
Floquet bands of KHM may be connected with the regular-to-chaos transition
in the classical limit.  Because ORDKR and KHM share the same classical limit
(up to a canonical transformation), we suspect that there should be some
similarity in their Floquet band topologies. Our second motivation for a
topological study is related to an earlier finding that, when $\hbar/(2\pi)$ is a \emph{rational} number, the spectral union of $U_{\text{ORDKR}-\alpha}$  (variant of ORDKR defined below)  over all $\alpha$ is the same as that of $U_{\text{KHM}-\alpha}$ (variant KHM defined below) over all $\alpha$ \cite{JMP}. This previous
mathematical result further suggests a possible topological connection
between the two models.   Interestingly, as we explore this possible topological connection,
we are able to see a connection between KHM propagator and ORDKR propagator for each individual value of $\alpha$ along with
an individual value of the Bloch phase,  thus going beyond Ref.~\cite{JMP} that considered a unification of all values of $\alpha$ and the Bloch phase. Further, as we shall see below, the connection is established by a mapping in the parameter space, which cannot be achieved by
a unitary transformation between the two propagators.

Next, we introduce necessary notation for our discussion of band topology.
To characterize the band topology for both ORDKR and KHM, we introduce
an additional periodic phase parameter $\alpha\in [0,2\pi)$ to the ORDKR
and KHM maps, namely,
\begin{equation}
\begin{split}
U_{\text{ORDKR}-\alpha}&=e^{i \frac{p^2}{2\hbar}}e^{-i \frac{K}{\hbar}\cos(q)}
e^{-i\frac{p^{2}}{2\hbar}}e^{-i\frac{L}{\hbar}\cos(q+\alpha)}\\
U_{\text{KHM}-\alpha}&=e^{-i\frac{L}{\hbar}\cos(p-\alpha)}e^{-i\frac{K}{\hbar}cos(q)}.
\end{split}
\end{equation}
For $\hbar=2\pi M/N$, both operators are periodic in (angular) momentum space with
period $N\hbar$. Hence, their eigenvalues are $2\pi$-periodic in the Bloch
phase $\varphi$ and also in $\alpha$, giving rise to $N$ \emph{extended}
Floquet bands which disperse as a function of $\varphi$ and $\alpha$. These
2-dimensional bands may be topologically characterized by Chern numbers,
denoted $C_n$ for the $n$th band. In what follows, we denote $|\psi_n(\varphi,\alpha)\rangle$
as an (generalized) eigenstate of  either $U_{\text{ORDKR}-\alpha}$ or $U_{\text{KHM}-\alpha}$,
in the $n$th band,
with an eigenvalue $\exp[ i \epsilon_n(\varphi,\alpha)]$.   Such a generalized
eigenstate lives on the entire (angular) momentum space.   We  then denote
$\tilde{U}(\varphi,\alpha)$ as the reduced $N\times N$ Floquet matrix
constructed from either $U_{\text{ORDKR}-\alpha}$ or $U_{\text{KHM}-\alpha}$
using the method described at the beginning of Section II.
We next define the state
$|\bar{\psi}_n(\varphi,\alpha)\rangle$, which is $|\psi_n(\varphi,\alpha)\rangle$
projected onto $N$ sites of one unit cell in the (angular) momentum space, i.e.,
$|\bar{\psi}_n(\varphi,\alpha)\rangle\equiv\sum_{m=0}^{N-1}\ket{m}\ip{m}{\psi_n(\varphi,\alpha)}$.
We further assume that $|\bar{\psi}_n(\varphi,\alpha)\rangle$ is normalized over one unit cell consisting
of $N$ sites.
Using the above notation, the Berry curvature of the $n$th band is then defined
as~\cite{derek12}
\begin{eqnarray}
B_{n}(\varphi,\alpha)=i\sum_{n'=1,\neq n}^{N}\Bigg\{\frac{
\langle\bar{\psi}_{n}|\frac{\partial\tilde{U}^{\dagger}}{\partial\varphi}|\bar{\psi}_{n'}\rangle
\langle\bar{\psi}_{n'}|\frac{\partial\tilde{U}}{\partial \alpha}|\bar{\psi}_{n}\rangle
}{|e^{-i\epsilon_{n}}-e^{-i\epsilon_{n'}}|^2}-\mathrm{c.c}\Bigg\},
\label{Bcurvature}
\end{eqnarray}
where we have suppressed the explicit dependences on $\varphi$ and $\alpha$
for brevity. From the Berry curvature we obtain the Chern number $C_n$,
\begin{eqnarray}
C_{n}=\frac{1}{2\pi}\int_{0}^{2\pi}d\varphi\int_{0}^{2\pi}d\alpha\:B_{n}(\varphi,\alpha).
\label{cherneq}
\end{eqnarray}


\subsection{Numerical Findings}

\begin{figure}[h!]
\begin{center}
$\begin{array}{c}
\includegraphics[trim=1cm 6.5cm 1cm 5cm,clip=true,height=!,width=14cm]{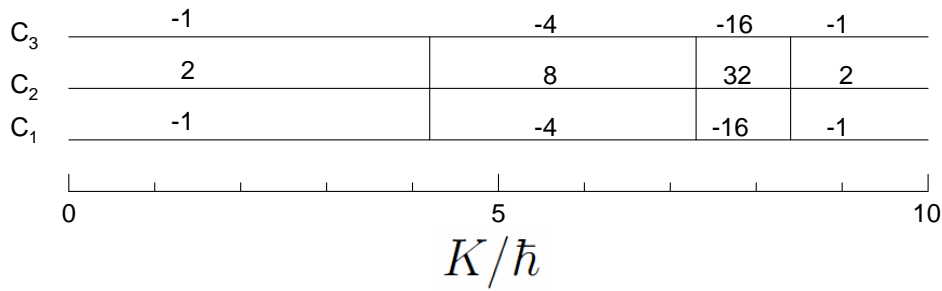}\\\
\end{array}$
\end{center}
\caption{Chern Numbers $C_n$ for \emph{both} ORDKR and KHM, for $K=L$.
In both cases, topological phase transitions occur at $K/\hbar\approx 4.20,7.25,8.40$
(correct to within $\pm 0.05$). }
\label{Cherntable_KeqL}
\end{figure}

We have conducted extensive numerical evaluations of the Floquet band Chern
numbers associated with both
$U_{\text{ORDKR}-\alpha}$ and $U_{\text{KHM}-\alpha}$. We find that for the
same $K$ and $L$ respectively in both models, the Chern numbers are always
equal. For example, for $\hbar=2\pi/3$ and $K=L$, Fig.~\ref{Cherntable_KeqL}
represents the Floquet band Chern numbers for \emph{both} models versus a
varying $K$. The Chern numbers obtained for $U_{\text{ORDKR}-\alpha}$ are
identical with those for $U_{\text{KHM}-\alpha}$. Here, we adopt the convention
that the band with largest absolute value of Chern number is always represented
by the line in the middle. Vertical lines represent collisions between quasi-energy
bands, during which Chern number transitions can take place. Note that in some
cases band 1 and band 3 can collide directly with each other through the boundary
of the quasienergy Brillouin zone.  It is also important to stress that the Chern
numbers of ORDKR match those of KHM for all $K$ values, despite their jumps
at various topological phase transition points.  We are thus clearly witnessing,
albeit numerically, a remarkable topological equivalence between ORDKR and KHM!



\begin{figure}[h!]
\begin{center}$
\begin{array}{cc}
\includegraphics[width=80mm, height=!]{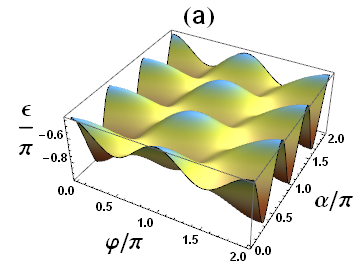}&
\includegraphics[width=80mm, height=!]{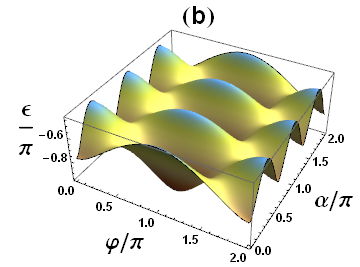}\\
\includegraphics[width=80mm, height=!]{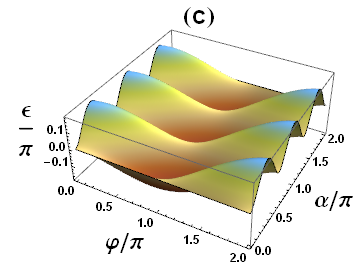}&
\includegraphics[width=80mm, height=!]{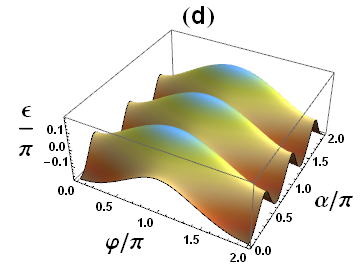}\\
\includegraphics[width=80mm, height=!]{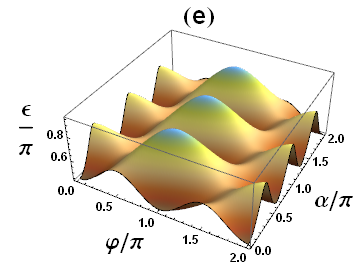}&
\includegraphics[width=80mm, height=!]{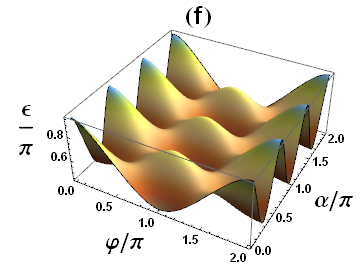}\\
\end{array}$
\end{center}
\caption{ (color online) Floquet band plots showing the quasienergy (eigenphase) dependence
on $\varphi$ and $\alpha$ in ORDKR and KHM with $K=L=3\hbar$,
$\hbar=2\pi/3$. Figs. (a),(c),(e) [(b),(d),(f)] belong to bands 1,2 and 3 respectively
for the ORDKR [KHM]. The ORDKR band profile appears to be a result of
some translation along the $\varphi$ and $\alpha$ axes followed by a rotation
of the spectrum about the $\epsilon$ axis.}
\label{KeqLplots}
\end{figure}

Some insight into this observed topological equivalence may be obtained by
comparing the quasienergy dispersions of the two models. In Fig.~\ref{KeqLplots},
we present the Floquet band structure for both ORDKR and KHM, in the case of
$K=L=3\hbar$. Interestingly, the ORDKR band profile appears to be the same as
that of KHM, up to some translation along the $\varphi$ and $\alpha$ axes,
followed by a rotation of the spectrum about the quasi-energy axis. This observation
is consistent with our proof of topological equivalence in the next section.


\begin{figure}[h!]
\begin{center}$
\begin{array}{c}
\includegraphics[trim=1cm 6.5cm 1cm 5cm,clip=true,height=!,width=14cm]{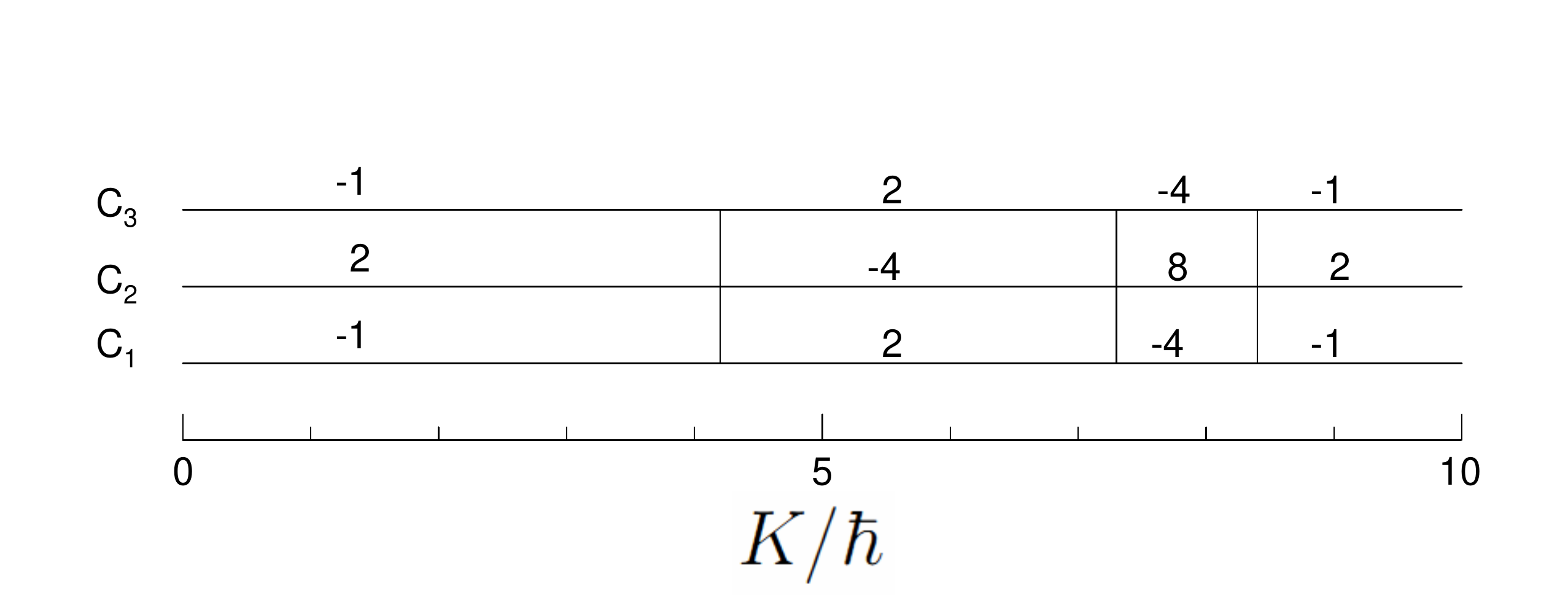}\\
\end{array}$
\end{center}
\caption{Chern Numbers $C_n$ for \emph{both} ORDKR and KHM, with
$\hbar=2\pi/3$, $L=\hbar$ fixed, and a varying $K$. In both cases, topological
phase transitions occur at $K/\hbar \approx 4.20, 7.25, 8.40$ (correct to within
$\pm 0.05$). The Chern numbers obtained here are different from the case of $K=L$ over some
ranges of $K$. Note that the phase transition points seem to be exactly the same
as those in Fig.~\ref{Cherntable_KeqL} only because we have rounded the phase transition points
to steps of 0.05.  A more accurate characterization does show very small differences.}
\label{Cherntable_KeqmL}
\end{figure}

We have numerically observed that the topological equivalence also occurs for
$K\ne L$. As one example of this, Fig.~\ref{Cherntable_KeqmL} depicts a zoo
of Chern numbers for ORDKR and KHM, with $\hbar=2\pi/3$, $L=\hbar$ fixed
but $K$ varying. We again see the same equivalence of Chern numbers across
a few topological phase transition points. In addition, we found computationally
that the Chern numbers are invariant upon an exchange between $L$ and $K$.
This was found to hold true also in other cases with more bands.


\begin{figure}[h!]
\begin{center}$
\begin{array}{cc}
\includegraphics[width=80mm, height=!]{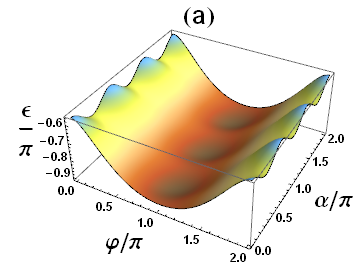}&
\includegraphics[width=80mm, height=!]{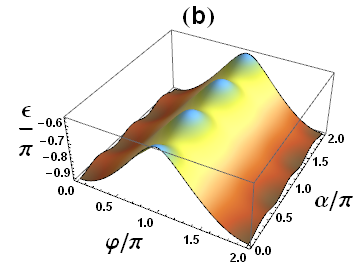}\\
\includegraphics[width=80mm, height=!]{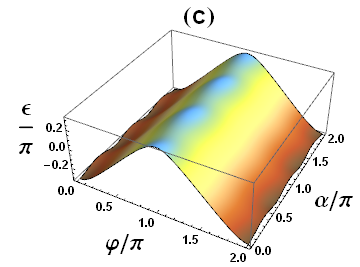}&
\includegraphics[width=80mm, height=!]{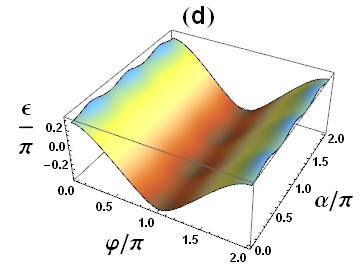}\\
\includegraphics[width=80mm, height=!]{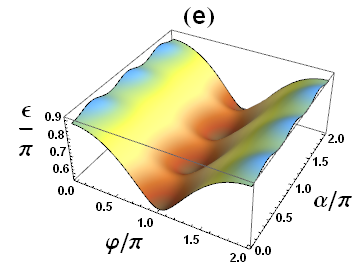}&
\includegraphics[width=80mm, height=!]{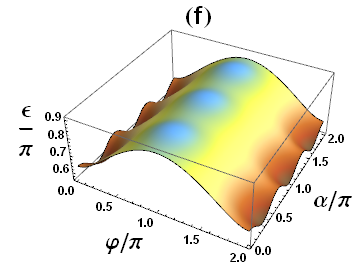}\\
\end{array}$
\end{center}
\caption{(color online) Floquet band plots showing the quasienergy (eigenphase) dependence
on $\varphi$ and $\alpha$, for $\text{ORDKR}$ and $\text{KHM}$ with
$K=3\hbar$,$L=\hbar$, $\hbar=2\pi/3$. Figures (a),(c),(e) [(b),(d),(f)] belong to
bands 1,2 and 3 respectively for the $\text{ORDKR}$ [$\text{KHM}$].}
\label{K_gt_Lplots}
\end{figure}

We have also plotted the Floquet band structure for a $K > L$ case in
Fig.~\ref{K_gt_Lplots}. Here we consider the case of $K/ \hbar=3$, $L/ \hbar=1$.
It is seen that the band profiles of ORDKR and KHM are once again similar and
appear to be related by a rotation and translation.


\subsection{Proof of Topological Equivalence}

To strictly confirm our claim of topological equivalence, we present an analytical
proof in this subsection. The proof proceeds as follows. We first show that the
reduced ORDKR Floquet matrix and the reduced KHM Floquet matrix are
equivalent up to a series of unitary transformations and a mapping between their
parameter values. We then show that these matrices obtained under the unitary
transformations and mapping of parameters still correspond to the same Chern
numbers as the original reduced matrices. These steps constitute a proof of
topological equivalence.


We consider cases with $\hbar=2\pi M/N$, with $M$ and $N$ co-prime and both
odd. In these cases, the reduced Floquet matrices of $U_{\text{ORDKR}-\alpha}$
and $U_{\text{KHM}-\alpha}$ (see the Appendix for details) can be written
compactly as a product of $N \times N$ unitary matrices
\begin{equation}
\begin{split}
{\tilde U}_{\text{ORDKR}}(\varphi , \alpha)
&=D_{\varphi}^\dagger D_1^\dagger\left(F^\dagger D_{1K}F\right)
D_1\left(F^\dagger D_{1L}F\right)D_{\varphi}\\
{\tilde U}_{\text{KHM}}(\varphi , \alpha)
&=D_{\varphi}^\dagger D_{2L}\left(F^\dagger D_{2K}F\right)D_{\varphi},
\end{split}
\end{equation}
where $D_{1K}$, $D_{1L}$, $D_{2K}$, $D_{2L}$ are diagonal unitary matrices,
with matrix elements
$(D_{1K})_{n,m}=e^{i\frac{K}{\hbar}\cos(\frac{2\pi}{N}n-\frac{\varphi}{N})}\delta _{n,m}$,
$(D_{1L})_{n,m}=e^{-i\frac{L}{\hbar}\cos(\frac{2\pi}{N}n-\frac{\varphi}{N}+\alpha)}\delta _{n,m}$,
$(D_{2K})_{n,m}=e^{-i\frac{K}{\hbar}\cos(\frac{2\pi}{N}n-\frac{\varphi}{N})}\delta _{n,m}$,
$(D_{2L})_{n,m}=e^{-i\frac{L}{\hbar}\cos(n\hbar-\alpha)}\delta _{n,m}$,
where the index $n$ takes values $0,1,\cdots,N-1$. $D_1$ and $D_\varphi $ are
defined as they were in Section II.

We begin the proof by applying a unitary transformation given by
$U_1 \equiv F^\dagger D_{2K}FD_{\varphi}$ to the
${\tilde U}_{\text{KHM}}(\varphi,\alpha)$ matrix to obtain
${\tilde V}_{\text{KHM}}(\varphi,\alpha)\equiv U_1\tilde{U}_{\text{KHM}}(\varphi,\alpha)U_1^\dagger$.
Writing $F^\dagger D_{2K} F$ as the exponential of a matrix, we obtain
\begin{equation}
\begin{split}
{\tilde V}_{\text{KHM}}(\varphi,\alpha)&=F^\dagger D_{2K}FD_{2L}\\
&=\exp\left[-i\frac{K}{2\hbar}F^\dagger
\begin{pmatrix}
\ddots 		& 			&\\
&e^{i(\frac{2\pi}{N}n-\frac{\varphi}{N})}+e^{-i(\frac{2\pi}{N}n-\frac{\varphi}{N})}&\\
			&			&\ddots\\
\end{pmatrix}
F\right]D_{2L}\\
&=\exp\left[-i\frac{K}{2\hbar}\left(e^{-i\frac{\varphi}{N}}C+
e^{i\frac{\varphi}{N}}C^\dagger\right)\right]
\begin{pmatrix}
\ddots 		& 			&\\
&e^{-i\frac{L}{\hbar}\cos(2\pi\frac{M}{N}n-\alpha)}&\\
			&			&\ddots\\
\end{pmatrix},
\end{split}
\label{unitaryKHM}
\end{equation}
where
\begin{equation}
C=
\begin{pmatrix}
0		&0		&\cdots	&0 	&1\\
1		&0		&\cdots	&0 	&0\\
0		&1		&\cdots	&0 	&0\\
\vdots	&\vdots	&\ddots	&\vdots 	&\vdots\\
0		&0		&\cdots	&1 	&0\\
\end{pmatrix}.
\end{equation}
In the following steps, we will apply a series of unitary transformations to the
reduced matrix $\tilde{U}_{\text{ORDKR}}(\varphi , \alpha)$ and show that the
result is equivalent to the above unitarily transformed version of
$\tilde{U}_{\text{KHM}}(\varphi , \alpha)$ provided a condition between
$\varphi$ and $\alpha$ in the two models is obeyed.

Applying a transformation given by $FD_{\varphi}$ to
$\tilde{U}_{\text{ORDKR}}(\varphi,\alpha)$, we obtain
$\tilde{U}^{(1)}_{\text{ORDKR}}(\varphi,\alpha)\equiv FD_{\varphi}
\tilde{U}_{\text{ORDKR}}(\varphi,\alpha)D_{\varphi}^\dagger F^\dagger$,
which we simplify as follows.
\begin{equation}
\begin{split}
{\tilde U}^{(1)}_{\text{ORDKR}}(\varphi,\alpha)
&=FD_1^\dagger F^\dagger D_{1K}FD_1F^\dagger D_{1L}\\
&=FD_1^\dagger\exp\left[i\frac{K}{2\hbar}\left(e^{-i\frac{\varphi}{N}}C+
e^{i\frac{\varphi}{N}}C^\dagger\right)\right]D_1F^\dagger D_{1L}\\
&=\exp\left[i\frac{K}{2\hbar}\left(e^{-i\frac{\varphi}{N}}FD_1^\dagger CD_1F^\dagger
+e^{i\frac{\varphi}{N}}FD_1^\dagger C^\dagger D_1F^\dagger\right)\right]D_{1L}.
\end{split}
\end{equation}
Denoting $X=FD_1^\dagger C D_1F^\dagger$,
${\tilde U}^{(1)}_{\text{ORDKR}}(\varphi,\alpha)=
\exp\left[i\frac{K}{2\hbar}\left(e^{-i\frac{\varphi}{N}}X+
e^{i\frac{\varphi}{N}}X^\dagger\right)\right]D_{1L}$.
The explicit expression for $X$ is
\begin{equation}
X=e^{i\pi\frac{N-M}{N}}
\begin{pmatrix}
	&	&	&e^{i\frac{2\pi}{N}\times M}	&\cdots	&0\\
	&	&	&\vdots					&\ddots	&\vdots\\
	&	&	&0	&\cdots	&e^{i\frac{2\pi}{N}\times (N-1)}\\
e^{i\frac{2\pi}{N}\times 0}&\cdots&0	& 	&		&\\
\vdots	&\ddots	&\vdots		&		& 		&\\
0&\cdots&e^{i\frac{2\pi}{N}\times (M-1)}&	&		&\\
\end{pmatrix}.
\end{equation}
Next, we introduce the $ N \times N $ permutation matrix $P_{\sigma}$ which is
made up entirely of zeroes except that in the $j$-th row, the $\sigma_{j}$-th column
equals 1, with $\sigma_{j}=j\times(N-M)\mod N$. Here, $j$ and $\sigma_{j}$ take
values $ 0, \cdots , N-1$. Note that $P_{\sigma}$ is unitary and that the set of
$\sigma_{j}$ values will include all of the $N$ values $j=0,1,\cdots,N-1$.
We apply the unitary transformation $P_{\sigma}$ to
${\tilde U}^{(1)}_{\text{ORDKR}}(\varphi,\alpha)$ and obtain
${\tilde U}^{(2)}_{\text{ORDKR}}(\varphi,\alpha)\equiv P_{\sigma}
{\tilde U}^{(1)}_{\text{ORDKR}}(\alpha, \varphi)P_{\sigma}^\dagger=
\exp\left[i\frac{K}{2\hbar}\left(e^{-i\frac{\varphi}{N}}P_{\sigma} XP_{\sigma}^\dagger
+e^{i\frac{\varphi}{N}}P_{\sigma} X^\dagger P_{\sigma}^\dagger\right)\right]D'_{1L}$,
where $ D'_{1L} \equiv P_{\sigma} D_{1L}P_{\sigma}^\dagger $.
$D'_{1L}$ is a diagonal unitary matrix with diagonal elements
$(D'_{1L})_{n,n}=e^{-i\frac{L}{\hbar}\cos(\frac{2\pi}{N}\sigma_n-\frac{\varphi}{N}
+\alpha)}=e^{-i\frac{L}{\hbar}\cos(-2\pi\frac{M}{N}n-\frac{\varphi}{N}+\alpha)}$.
The effect of the permutation matrix on $X$ is as follows.
\begin{equation}
P_{\sigma} X P_{\sigma}^\dagger=e^{i\pi\frac{N-M}{N}}
\begin{pmatrix}
0		&0		&\cdots	&0		&e^{i\frac{2\pi}{N} \sigma_{N-1}}\\
e^{i\frac{2\pi}{N}\sigma_0}&0&\cdots  &0			&0\\
0	&e^{i\frac{2\pi}{N}\sigma_1}	&	&0			&0\\
\vdots 	&\vdots		&\ddots		&\vdots		&\vdots\\
0		&0	&\cdots	&e^{i\frac{2\pi}{N}\sigma_{N-2}}	&0\\
\end{pmatrix}  
\end{equation}
We can see that the structure of the above matrix is very similar to $C$ and would be made
identical with it if we were to replace all the nonzero elements with 1. This is
achieved by a transformation via the diagonal unitary matrix $D_0$ which
has diagonal elements
$ (D_0)_{n,n}=e^{-i \left[\frac{2\pi}{N}\sum_{k=0}^{k=n-1}\sigma_k+\pi\frac{N-M}{N}n\right]}$.
It can be shown that $D_0 P_{\sigma} X P_{\sigma}^\dagger D_0 ^\dagger = C$.
Denoting
${\tilde V}_{\text{ORDKR}}(\varphi,\alpha)\equiv D_0
{\tilde U}^{(2)}_{\text{ORDKR}}(\varphi,\alpha) D_0^\dagger $ and using that
$ D_0$ and $ D'_{1L}$ commute due to their both being diagonal, we obtain
\begin{equation}
\begin{split}
{\tilde V}_{\text{ORDKR}}(\varphi,\alpha)
&=\exp\left[i\frac{K}{2\hbar}\left(e^{-i\frac{\varphi}{N}}C
+e^{i\frac{\varphi}{N}}C^\dagger \right)\right]D'_{1L}\\
&=\exp\left[-i\frac{K}{2\hbar}\left(e^{-i\frac{\varphi+N\pi}{N}}C
+e^{i\frac{\varphi+N\pi}{N}}C^\dagger \right)\right]
\begin{pmatrix}
\ddots 		& 			&\\
&e^{-i\frac{L}{\hbar}\cos(2\pi\frac{M}{N}j+\frac{\varphi}{N}-\alpha)}&\\
			&			&\ddots\\
\end{pmatrix}.
\end{split}
\label{unitaryORDKR}
\end{equation}


From Eq.~(\ref{unitaryKHM}) and (\ref{unitaryORDKR}), we observe that
$ {\tilde V}_{\text{ORDKR}}(\varphi,\alpha)$ and
${\tilde V}_{\text{KHM}}(\tilde{\varphi},\tilde{\alpha})$ are identical, provided that
$\tilde\varphi=\varphi+N\pi$ and
$\tilde\alpha=\alpha-\frac{\varphi}{N}$.
Summarizing what we have found so far, we have learned that if we unitarily
transform from
${\tilde U}_{\text{KHM}}(\tilde{\varphi},\tilde{\alpha})$ to
$ {\tilde V}_{\text{KHM}}(\tilde{\varphi},\tilde{\alpha})\equiv U_1
{\tilde U}_{\text{KHM}}(\tilde{\varphi},\tilde{\alpha})U_1^\dagger$, where
$U_1\equiv F^\dagger D_{2K}FD_{\tilde{\varphi}}$, and unitarily transform from
${\tilde U}_{\text{ORDKR}}(\varphi,\alpha)$ to
${\tilde V}_{\text{ORDKR}}(\varphi,\alpha)\equiv U_2
{\tilde U}_{\text{ORDKR}}(\varphi,\alpha)U_2^\dagger$, where
$U_2\equiv D_0P_{\sigma}FD_{\varphi}$,
we find that the two unitarily transformed matrices are identical up to some
mapping between $(\tilde{\varphi},\tilde{\alpha})$ and $(\varphi,\alpha)$.

\begin{figure}[h!]
\begin{center}
$\begin{array}{c}
\includegraphics[trim=0cm 0cm 0cm 0cm,clip=true,height=18cm ,width=!]{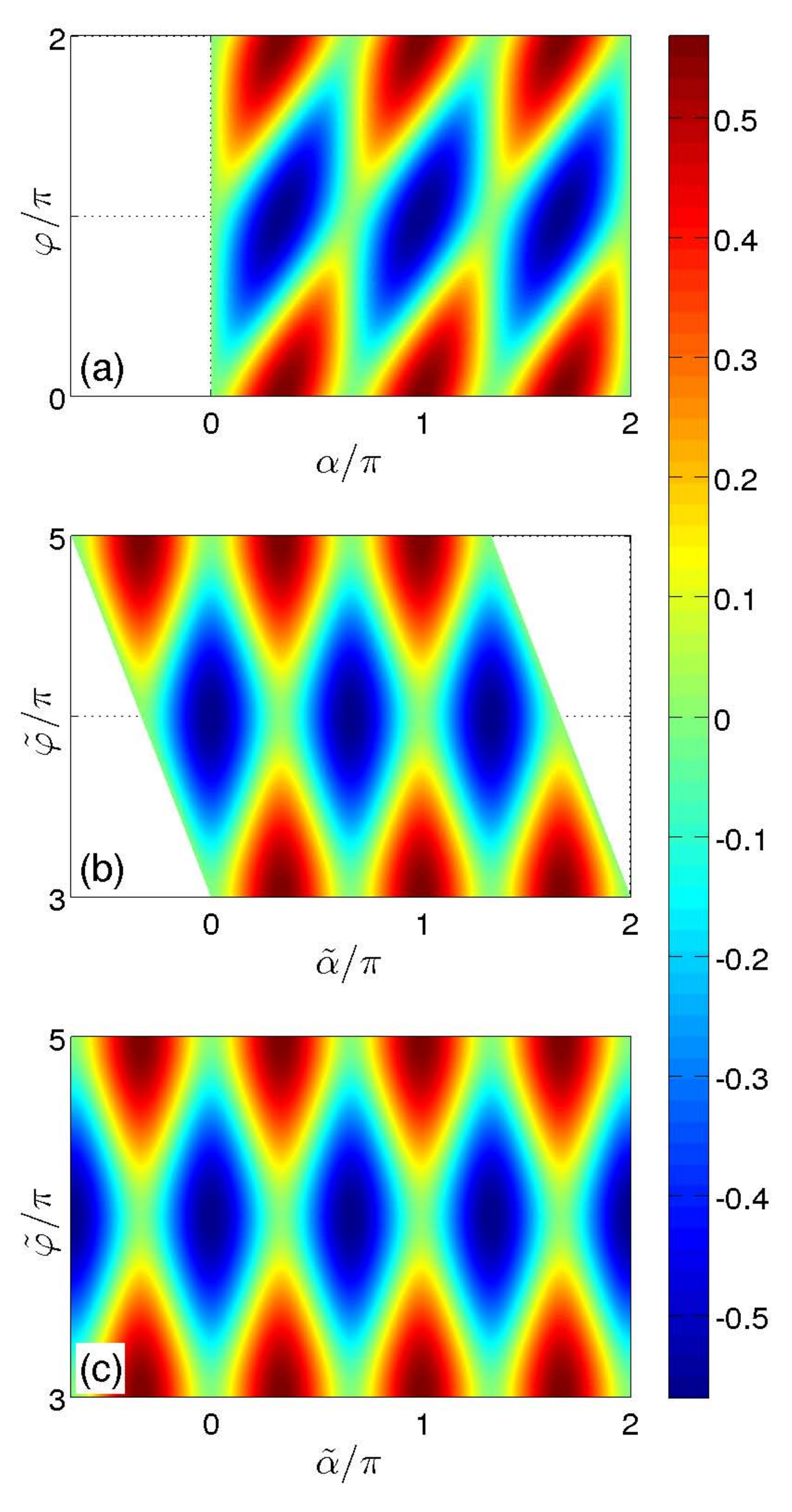}\\
\end{array}$
\end{center}
\caption{ (color online) Quasi-energy band (band 2) plot for $K=L=3\hbar$ with $\hbar=2\pi/3$.
Panel (a) shows dependence on ($\varphi$, $\alpha$) for ${\tilde U}_\text{ORDKR}(\varphi,\alpha)$, whereas panel (b) shows dependence on ($\tilde\varphi$, $\tilde\alpha$) for ${\tilde U}_\text{ORDKR}(\tilde\varphi-N\pi,\tilde\alpha+\frac{\tilde\varphi}{N}-\pi)$.
Panel (c) shows dependence on ($\tilde\varphi$, $\tilde\alpha$) for ${\tilde U}_\text{KHM}(\tilde\varphi,\tilde\alpha)$.}
\label{Spectrun_compare}
\end{figure}

Figure~\ref{Spectrun_compare} represents one example of the quasi-energy band plot
for both ORDKR and KHM. Referring to panel (b) and panel (c), we thus directly see that provided that
$\tilde\varphi=\varphi+N\pi$ and
$\tilde\alpha=\alpha-\frac{\varphi}{N}$, the extended Floquet band structure for
ORDKR and KHM are the same (though the boundaries on the $(\tilde\varphi,\tilde\alpha)$ plane are different).

Recapping our proof so far,  with the mapping $\tilde\varphi =\varphi+N\pi$ and
$\tilde\alpha =\alpha-\frac{\varphi}{N}$,  we have the following
\begin{equation}
{\tilde U}_{\text{KHM}}(\tilde{\varphi},\tilde{\alpha})= U_{T}{\tilde U}_{\text{ORDKR}}(\varphi,\alpha)U_T^\dagger,
\label{eqn:equvilance}
\end{equation}
where $U_T\equiv D_{\tilde{\varphi}}^\dagger F^\dagger D_{2K}^\dagger(\tilde{\varphi})FD_0P_{\sigma}FD_{\varphi}$, and the definitions of the matrices $D$, $F$, $D_{2K}$, $D_0$ and $P_{\sigma}$ all are previously given. For example,
$(D_{\tilde{\varphi}})_{n,m}=e^{-in\frac{\tilde{\varphi}}{N}}\delta _{n,m}$,
and $(D_{2K})(\tilde{\varphi})_{n,m}=e^{-i\frac{K}{\hbar}\cos(\frac{2\pi}{N}n-\frac{\tilde{\varphi}}{N})}\delta _{n,m}$.
Let $|\bar{\psi}_n^{\text{KHM}}(\tilde{\varphi},\tilde{\alpha})\rangle$ be the $n$th eigenstate of
${\tilde U}_{\text{KHM}}(\tilde{\varphi},\tilde{\alpha})$ and
$|\bar{\psi}_n^{\text{ORDKR}}(\varphi,\alpha)\rangle$ be the $n$th eigenstate of  ${\tilde U}_{\text{ORDKR}}(\varphi,\alpha)$.
Equation (\ref{eqn:equvilance}) then leads to
\begin{equation}
|\bar{\psi}_n^{\text{KHM}}(\tilde{\varphi},\tilde{\alpha})\rangle =U_T |\bar{\psi}_n^{\text{ORDKR}}({\varphi},{\alpha})\rangle.
\end{equation}
Because scanning all the values of $(\varphi,\alpha)$ will scan all the values of $(\tilde{\varphi},\tilde{\alpha})$, it is obvious now that the union of the spectrum of ${\tilde U}_{\text{KHM}}(\tilde{\varphi},\tilde{\alpha})$ (after considering all values of $\tilde{\varphi}$ and $\tilde{\alpha}$)  should be the same
as the union of the spectrum of ${\tilde U}_{\text{ORDKR}}(\varphi,\alpha)$ (after considering all values of $\varphi$ and $\alpha$), thus directly confirming an early proof in Ref.~\cite{JMP}.  We stress, however, that the one-to-one correspondence between
$\tilde{U}_{\text{ORDKR}}(\varphi,\alpha)$ and $\tilde{U}_{\text{KHM}}(\varphi,\alpha)$ is a new result that we did not find previously.

Finally, we show that $ {\tilde V}_{\text{ORDKR}}(\varphi,\alpha)$ and
${\tilde V}_{\text{KHM}}(\varphi,\alpha)$ have the same set of Chern numbers as
their respective original matrices ${\tilde U}_{\text{ORDKR}}(\varphi,\alpha)$ and
${\tilde U}_{\text{KHM}}(\varphi,\alpha)$. To do this, we make use of the line
integral version of the Chern number of the $n$th band given by
\begin{equation}
C_n=\frac{i}{2\pi}\oint d\vec{\theta}\bra{\bar{\psi} _{n}(\vec{\theta})}
\partial _{\vec{\theta}} \ket{\bar{\psi} _{n} (\vec{\theta})},
\label{lineint}
\end{equation}
where $\vec{\theta} \equiv (\varphi,\alpha) $ and the line integral is around the
perimeter of the Brillouin zone $ (0,2\pi] \times (0,2 \pi]$ in $(\varphi,\alpha) $
parameter space. Here $ \ket{\bar{\psi} _{n}(\vec{\theta})} $ again refers to the $n$th
band eigenstate of either ${\tilde U}_{\text{ORDKR}}(\varphi , \alpha)$ or
${\tilde U}_{\text{KHM}}(\varphi , \alpha)$ at the point $\vec{\theta}$. The
eigenstates of $ {\tilde V}_{\text{KHM}}(\varphi , \alpha)$ and
${\tilde V}_{\text{ORDKR}}(\varphi,\alpha)$, denoted
$\ket{\tilde{\psi}_{n}(\vec{\theta})}$, are related to the original eigenstates by
$U_{1,2}^\dagger\ket{\tilde{\psi}_{n}(\vec{\theta})}=\ket{\bar{\psi} _{n}(\vec{\theta})}$
respectively. We may substitute this into Eq.~(\ref{lineint}) and obtain an expression
for $C_n$ in terms of $\ket{\tilde{\psi}_{n}(\vec{\theta})}$. Because the transformations
$U_{1,2}$ depend on $\varphi$ but not on $\alpha$, it can be shown, by making
use of the fact that the line integrals along $\alpha=0$ and $\alpha=2\pi$ are in
opposite directions, that the resulting expression for $C_n$ reduces back to that
of the form of Eq.~(\ref{lineint}), except with the transformed eigenstates taking
the place of the original ones. This proves that the Chern numbers of the unitarily
transformed reduced matrices match those of the original ones.

Next, we note that when we impose $\tilde \varphi=\varphi+N\pi$ and
$\tilde \alpha=\alpha-\frac{\varphi}{N}$, working out the line integral in
Eq.~(\ref{lineint}) for ${\tilde V}_{\text{ORDKR}}(\varphi,\alpha)$ over a typical
square perimeter space in $(\varphi,\alpha)$ space with corners
$(0,0), (2\pi,0),(2\pi,2\pi),(0,2\pi)$ is equivalent to working out the line integral
for ${\tilde V}_{\text{KHM}}(\tilde{\varphi},\tilde{\alpha})$ over some parallelogram
in $(\tilde{\varphi},\tilde{\alpha})$ space with corners
$(N\pi,0),( N\pi+2\pi,-2\pi/N),(N\pi+2\pi,2\pi-2\pi/N),(N\pi,2\pi)$. To complete the
proof of topological equivalence, we need only show that the aforementioned
line integral in $(\tilde{\varphi},\tilde{\alpha})$ for
${\tilde V}_{\text{KHM}}(\tilde{\varphi},\tilde{\alpha}) $ gives a result equal to
that when we calculate the line integral around the perimeter of the usual
$(0,2\pi]\times(0,2\pi]$ Brillouin zone. However, this can be easily shown to be
the case by converting the line integral around the parallelogram into a surface
integral using Stokes' theorem. We then obtain a surface integral of the form of
Eq.~(\ref{cherneq}) enclosing the area of the parallelogram. Because the Berry
curvature as seen in Eq.~(\ref{Bcurvature}) is exactly $2\pi$-periodic along both
$\varphi$ and $\alpha$, it is trivial to see that we can map the area of the
parallelogram back onto that of the original $ (0,2\pi] \times (0,2 \pi]$ Brillouin
zone, without any difference in the result of the integral. In other words, the
Chern numbers of $ {\tilde V}_{\text{KHM}}(\tilde{\varphi} , \tilde{\alpha}) $ and
${\tilde V}_{\text{ORDKR}}(\varphi,\alpha)$ are always identical. Putting this
together with the result of the previous paragraph, we may conclude that the
Chern numbers of the original matrices ${\tilde U}_{\text{KHM}}(\varphi , \alpha)$
and  ${\tilde U}_{\text{ORDKR}}(\varphi , \alpha)$ are indeed the same. This
completes our proof of topological equivalence.

\section{Concluding Remarks}
In this work, we have mainly focused on two topics: the spectral difference
between ORDKR and KHM (comparing quantum maps $U_{\text{ORDKR}}$
and $U_{\text{KHM}}$) and their topological equivalence upon introducing
an additional periodic phase parameter $\alpha$ (comparing quantum maps
$U_{\text{ORDKR}-\alpha}$ and $U_{\text{KHM}-\alpha}$). One important
spectral difference we have found is the existence of a flat band for
$U_{\text{ORDKR}}$ {under the condition $K=L$}, but not for $U_{\text{KHM}}$. To our knowledge, this
is the first example of a periodically driven model that has a mixture of flat
band and non-flat bands. States launched from a flat band will be strictly
localized, and this feature might be useful for benchmarking experimental
errors in any future realizations of ORDKR.  The coexistence of a flat band with non-flat bands may
also open up new applications of $\delta$-kicked systems. We have also
shown that for small kick strength $K=L$, the band width of the non-flat
bands of $U_{\text{ORDKR}}$ scales with $K$ in a power law with a high
exponent $N+2$, indicating that for sufficiently small kick strength, all
Floquet bands will be effectively flat for a long time scale.  The dynamical
consequence is a transient dynamical localization in ORDKR (absent in KHM)
for a long time scale. The topological equivalence between
$U_{\text{ORDKR}-\alpha}$ and $U_{\text{KHM}-\alpha}$ makes our
ORDKR-KHM comparison even more interesting. That is, for a fixed $\alpha$,
ORDKR and KHM have many different features. But topologically speaking,
upon introducing one extra parameter $\alpha$ we have a topological equivalence
between an extended ORDKR model, previously proposed in studies of
quantum ratchet acceleration without using a bichromatic lattice \cite{PRE08},
with a simple extension of the standard KHM. To have a pair of models that
are topologically equivalent should be a useful  contribution to the general
understanding of the topological properties of periodically driven
systems~\cite{dana}.

\section{Acknowledgments}
J.W. and J.G. acknowledge helpful discussions with Prof. Italo Guarneri,
who also confirmed our flat-band result by showing us an alternative
proof by him. D.H. thanks Adam Zaman Chaudhry for helpful discussions.
J.W. received support from NNSF (Grants No.11275159 and No. 10925525) and SRFDP
(Grant No. 20100121110021) of China, and J.G. is supported
by ARF Tier I, MOE of Singapore (Grant No. R-144-000-276-112).
J.G. dedicates this work to his late beloved wife Huairui Zhang.

\appendix
\section{Expressions for reduced Floquet matrices}
\label{ReduceFloMat.apx}

For $\hbar=2\pi M/N$ with $M$ and $N$ being coprime and odd integers,
reduced $N\times N$ Floquet matrix is given by
$\left[{\tilde U}(\varphi)\right]_{n,m}=\sum_{l=-\infty}^{\infty}\bra{n}\hat U \ket{m+l\times N}e^{il\varphi}$.

\subsection{Reduced Floquet matrix for ORDKR}

The Floquet operator of ORDKR is
\begin{eqnarray}
{U}_{\text{ORDKR}}=e^{i\frac{p^2}{2\hbar}}e^{-i\frac{K}{\hbar}\cos(q)}
e^{-i\frac{p^2}{2\hbar}}e^{-i\frac{L}{\hbar}\cos(q)}.
\end{eqnarray}
The reduced $N\times N$ Floquet matrix is thus
\begin{eqnarray}
\begin{split}
\left[\tilde{U}_{\text{ORDKR}}(\varphi)\right]_{n, m}&=\sum_{l=-\infty}^{\infty}
\bra{n}\hat U_{\text{ORDKR}}\ket{m+l\times N} e^{il\varphi}\\
&=\sum_{l=-\infty}^{\infty}\sum_{l'=-\infty}^{\infty}\sum_{m'=0}^{N-1}
\bra{n}e^{i\frac{p^2}{2\hbar}}e^{-i\frac{K}{\hbar}\cos(q)}\ket{m'+l'\times N}\\
&\qquad\times \bra{m'+l'\times N}e^{-i\frac{p^2}{2\hbar}}
e^{-i\frac{L}{\hbar}\cos(q)}\ket{m+l\times N} e^{il\varphi}\\
&=\sum_{m'=0}^{N-1}\frac{1}{2\pi}e^{i\frac{\hbar}{2}n^2}\int_0^{2\pi} d\theta_2
e^{-i\frac{K}{\hbar}\cos(\theta_2)}e^{i\theta_2(m'-n)}\sum_{l'=-\infty}^{\infty}e^{i\theta_2l'N}\\
&\qquad\times\frac{1}{2\pi}e^{-i\frac{\hbar}{2}(m'+l'N)^2}\int_0^{2\pi} d\theta_1
e^{-i\frac{L}{\hbar}\cos(\theta_1)}e^{i \theta_1(m-m')}\sum_{l=-\infty}^{\infty}e^{i \theta_1(l-l')N}e^{il\varphi}\\
&=\sum_{m'=0}^{N-1}\frac{1}{2\pi} e^{i\frac{\hbar}{2}n^2}\int_0^{2\pi} d\theta_2
e^{i\frac{K}{\hbar}\cos(\theta_2+\pi)}e^{i(\theta_2+\pi)(m'-n)}e^{i\pi(m'-n)}\sum_{l'=-\infty}^{\infty}e^{i (\theta_2+\pi) l'N }\\
&\qquad\times\frac{1}{2\pi} e^{-i\frac{\hbar}{2}m'^2}\int_0^{2\pi} d\theta_1
e^{-i\frac{L}{\hbar}\cos(\theta_1)}e^{i\theta_1(m-m')}e^{-i \theta_1 l'N}\sum_{l=-\infty}^{\infty}e^{i \theta_1 lN }e^{il\varphi}.
\end{split}
\end{eqnarray}

To simplify, we make use of the Poisson summation formula
\begin{eqnarray}
\sum_{l=-\infty}^{\infty}e^{2\pi i l\varphi}=\sum_{j=-\infty}^{\infty}\delta(\varphi-j),
\end{eqnarray}
and obtain
\begin{eqnarray}
\begin{split}
\left[\tilde{U}_{\text{ORDKR}}(\varphi)\right]_{n, m}
&=e^{i\frac{\hbar}{2}n^2}\sum_{m'=0}^{N-1}e^{-i\frac{\hbar}{2}m'^2}e^{i \pi(m'-n)}
\frac{1}{N}\sum_{j_2=0}^{N-1}e^{ i\frac{K}{\hbar} \cos(\frac{2\pi}{N}j_2-\frac{\varphi}{N})}
e^{i (\frac{2\pi}{N}j_2-\frac{\varphi}{N})(m'-n)}\\
&\qquad\times\frac{1}{N}\sum_{j_1=0}^{N-1}e^{ -i \frac{L}{\hbar}
\cos(\frac{2\pi}{N}j_1-\frac{\varphi}{N})}e^{i (\frac{2\pi}{N}j_1-\frac{\varphi}{N})(m-m')}\\
&=\frac{1}{N^2}\sum_{j_2=0}^{N-1}\sum_{m'=0}^{N-1}\sum_{j_1=0}^{N-1}
e^{i n\frac{\varphi}{N}}e^{-i m\frac{\varphi}{N}}\\
&\qquad\times e^{-i\frac{2\pi-\hbar}{2}n^2}e^{-i \frac{2\pi}{N}n j_2}
e^{ i\frac{K}{\hbar} \cos(\frac{2\pi}{N}j_2-\frac{\varphi}{N})}e^{i \frac{2\pi}{N}j_2 m'}\\
&\qquad\times e^{i\frac{2\pi-\hbar}{2}m'^2}e^{-i \frac{2\pi}{N}m' j_1}
e^{ -i \frac{L}{\hbar} \cos(\frac{2\pi}{N}j_1-\frac{\varphi}{N})}e^{i \frac{2\pi}{N}j_1 m}.
\end{split}
\end{eqnarray}

For the sake of illustration, we write the reduced Floquet matrix as a product of unitary matrices
\begin{equation}
\begin{split}
&\tilde{U}_{\text{ORDKR}}(\varphi)=
\begin{pmatrix}
\ddots 		& 			&\\
	&e^{in\frac{\varphi}{N}}	&\\
			&			&\ddots\\
\end{pmatrix}\\
&
\begin{pmatrix}
\ddots 		& 			&\\
	&e^{-i\frac{2\pi-\hbar}{2}n^2}&\\
			&			&\ddots\\
\end{pmatrix}
\begin{pmatrix}
			& 			&\\
&\frac{e^{-i\frac{2\pi}{N}n j_2}}{\sqrt N} 	&\\
			&			&\\
\end{pmatrix}
\begin{pmatrix}
\ddots 		& 			&\\
&e^{i\frac{K}{\hbar}\cos(\frac{2\pi}{N}j_2-\frac{\varphi}{N})}	&\\
			&			&\ddots\\
\end{pmatrix}
\begin{pmatrix}
			& 			&\\
&\frac{e^{i\frac{2\pi}{N}j_2 m'}}{\sqrt N}	&\\
			&			&\\
\end{pmatrix}\\
&
\begin{pmatrix}
\ddots 		& 			&\\
	&e^{i\frac{2\pi-\hbar}{2}m'^2}&\\
			&			&\ddots\\
\end{pmatrix}
\begin{pmatrix}
			& 			&\\
&\frac{e^{-i\frac{2\pi}{N}m' j_1}}{\sqrt N} 	&\\
			&			&\\
\end{pmatrix}
\begin{pmatrix}
\ddots 		& 			&\\
&e^{-i\frac{L}{\hbar}\cos(\frac{2\pi}{N}j_1-\frac{\varphi}{N})}	&\\
			&			&\ddots\\
\end{pmatrix}
\begin{pmatrix}
			& 			&\\
&\frac{e^{i\frac{2\pi}{N}j_1 m}}{\sqrt N}	&\\
			&			&\\
\end{pmatrix}\\
&\qquad\qquad\qquad\begin{pmatrix}
\ddots 		& 			&\\
	&e^{-im\frac{\varphi}{N}}	&\\
			&			&\ddots\\
\end{pmatrix}\\
\end{split}
\end{equation}

If we introduce an additional periodic phase parameter $\alpha\in [0,2\pi)$ to the ORDKR map, the Floquet operator becomes
\begin{equation}
U_{\text{ORDKR}-\alpha}=e^{i \frac{p^2}{2\hbar}}e^{-i \frac{K}{\hbar}\cos(q)}e^{-i\frac{p^{2}}{2\hbar}}e^{-i\frac{L}{\hbar}\cos(q+\alpha)}.
\end{equation}

The corresponding reduced Floquet matrix is
\begin{eqnarray}
\begin{split}
\left[\tilde{U}_{\text{ORDKR}}(\varphi,\alpha)\right]_{n, m}
&=\frac{1}{N^2}\sum_{j_2=0}^{N-1}\sum_{m'=0}^{N-1}\sum_{j_1=0}^{N-1}e^{i n\frac{\varphi}{N}}e^{-i m\frac{\varphi}{N}}\\
&\qquad\times e^{-i\frac{2\pi-\hbar}{2}n^2}e^{-i \frac{2\pi}{N}n j_2}e^{ i\frac{K}{\hbar} \cos(\frac{2\pi}{N}j_2-\frac{\varphi}{N})}e^{i \frac{2\pi}{N}j_2 m'}\\
&\qquad\times e^{i\frac{2\pi-\hbar}{2}m'^2}e^{-i \frac{2\pi}{N}m' j_1}e^{ -i \frac{L}{\hbar} \cos(\frac{2\pi}{N}j_1-\frac{\varphi}{N}+\alpha)}e^{i \frac{2\pi}{N}j_1 m}.
\end{split}
\end{eqnarray}
Written again as a product of unitary matrices,
\begin{equation}
\begin{split}
&\tilde{U}_{\text{ORDKR}}(\varphi,\alpha)=
\begin{pmatrix}
\ddots 		& 			&\\
	&e^{in\frac{\varphi}{N}}	&\\
			&			&\ddots\\
\end{pmatrix}\\
&
\begin{pmatrix}
\ddots 		& 			&\\
	&e^{-i\frac{2\pi-\hbar}{2}n^2}&\\
			&			&\ddots\\
\end{pmatrix}
\begin{pmatrix}
			& 			&\\
&\frac{e^{-i\frac{2\pi}{N}n j_2}}{\sqrt N} 	&\\
			&			&\\
\end{pmatrix}
\begin{pmatrix}
\ddots 		& 			&\\
&e^{i\frac{K}{\hbar}\cos(\frac{2\pi}{N}j_2-\frac{\varphi}{N})}	&\\
			&			&\ddots\\
\end{pmatrix}
\begin{pmatrix}
			& 			&\\
&\frac{e^{i\frac{2\pi}{N}j_2 m'}}{\sqrt N}	&\\
			&			&\\
\end{pmatrix}\\
&
\begin{pmatrix}
\ddots 		& 			&\\
	&e^{i\frac{2\pi-\hbar}{2}m'^2}&\\
			&			&\ddots\\
\end{pmatrix}
\begin{pmatrix}
			& 			&\\
&\frac{e^{-i\frac{2\pi}{N}m' j_1}}{\sqrt N} 	&\\
			&			&\\
\end{pmatrix}
\begin{pmatrix}
\ddots 		& 			&\\
&e^{-i\frac{L}{\hbar}\cos(\frac{2\pi}{N}j_1-\frac{\varphi}{N}+\alpha)}	&\\
			&			&\ddots\\
\end{pmatrix}
\begin{pmatrix}
			& 			&\\
&\frac{e^{i\frac{2\pi}{N}j_1 m}}{\sqrt N}	&\\
			&			&\\
\end{pmatrix}\\
&\qquad\qquad\qquad\begin{pmatrix}
\ddots 		& 			&\\
	&e^{-im\frac{\varphi}{N}}	&\\
			&			&\ddots\\
\end{pmatrix}\\
\end{split}
\end{equation}

\subsection{Reduced Floquet matrix for KHM}

The Floquet operator of KHM is
\begin{eqnarray}
U_{\text{KHM}}=e^{-i\frac{L}{\hbar}\cos(p)}e^{-i\frac{K}{\hbar}\cos(q)},
\end{eqnarray}
with reduced $N\times N$ Floquet matrix
\begin{eqnarray}
\begin{split}
\left[\tilde{U}_{\text{KHM}}(\varphi)\right]_{n, m}&=\sum_{l=-\infty}^{\infty}\langle n|\hat U_{\text{KHM}}|m+l\times N\rangle e^{il\varphi}\\
&=\frac{1}{2\pi}e^{-i\frac{L}{\hbar}\cos(n\hbar)}\int_0^{2\pi} d\theta e^{-i\frac{K}{\hbar}  \cos(\theta)}e^{i \theta(m-n)}\sum_{l=-\infty}^{\infty}e^{i \theta lN } e^{il\varphi}\\
&=e^{-i\frac{L}{\hbar}\cos(n\hbar)}\frac{1}{N}\sum_{j=0}^{N-1}e^{-i\frac{K}{\hbar} \cos(\frac{2\pi}{N}j-\frac{\varphi}{N})}e^{i (\frac{2\pi}{N}j-\frac{\varphi}{N})(m-n)}\\
&=\frac{1}{N}\sum_{j=0}^{N-1}e^{i n\frac{\varphi}{N}}e^{-i\frac{L}{\hbar}\cos(n\hbar)}e^{-i \frac{2\pi}{N}n j}e^{ i\frac{K}{\hbar} \cos(\frac{2\pi}{N}j-\frac{\varphi}{N})}e^{i \frac{2\pi}{N}jm}e^{-i m\frac{\varphi}{N}}.
\end{split}
\end{eqnarray}

For the sake of illustration, we write the reduced Floquet matrix as a product of unitary matrices.
\begin{equation}
\begin{split}
&\tilde{U}_{\text{KHM}}(\varphi)=
\begin{pmatrix}
\ddots 		& 			&\\
	&e^{in\frac{\varphi}{N}}	&\\
			&			&\ddots\\
\end{pmatrix}\\
&
\begin{pmatrix}
\ddots 		& 			&\\
	&e^{-i\frac{L}{\hbar}\cos(n\hbar)}&\\
			&			&\ddots\\
\end{pmatrix}
\begin{pmatrix}
			& 			&\\
&\frac{e^{-i\frac{2\pi}{N}n j}}{\sqrt N} 	&\\
			&			&\\
\end{pmatrix}
\begin{pmatrix}
\ddots 		& 			&\\
&e^{-i\frac{K}{\hbar}\cos(\frac{2\pi}{N}j-\frac{\varphi}{N})}	&\\
			&			&\ddots\\
\end{pmatrix}
\begin{pmatrix}
			& 			&\\
&\frac{e^{i\frac{2\pi}{N}j m}}{\sqrt N}	&\\
			&			&\\
\end{pmatrix}\\
&\qquad\qquad\qquad\begin{pmatrix}
\ddots 		& 			&\\
	&e^{-im\frac{\varphi}{N}}	&\\
			&			&\ddots\\
\end{pmatrix}\\
\end{split}
\end{equation}

If we introduce an additional periodic phase parameter $\alpha\in [0,2\pi)$ to the KHM map, the Floquet operator becomes
\begin{equation}
U_{\text{KHM}-\alpha}=e^{-i\frac{L}{\hbar}\cos(p-\alpha)}e^{-i\frac{K}{\hbar}cos(q)}.
\end{equation}

The corresponding reduced Floquet matrix is
\begin{eqnarray}
\left[\tilde{U}_{\text{KHM}}(\varphi,\alpha)\right]_{n, m}
=\frac{1}{N}\sum_{j=0}^{N-1}e^{i n\frac{\varphi}{N}}e^{-i\frac{L}{\hbar}\cos(n\hbar-\alpha)}e^{-i \frac{2\pi}{N}n j}e^{ i\frac{K}{\hbar} \cos(\frac{2\pi}{N}j-\frac{\varphi}{N})}e^{i \frac{2\pi}{N}jm}e^{-i m\frac{\varphi}{N}}.
\end{eqnarray}
Written as a product of unitary matrices,
\begin{equation}
\begin{split}
&\tilde{U}_{\text{KHM}}(\varphi,\alpha)=
\begin{pmatrix}
\ddots 		& 			&\\
	&e^{in\frac{\varphi}{N}}	&\\
			&			&\ddots\\
\end{pmatrix}\\
&
\begin{pmatrix}
\ddots 		& 			&\\
	&e^{-i\frac{L}{\hbar}\cos(n\hbar-\alpha)}&\\
			&			&\ddots\\
\end{pmatrix}
\begin{pmatrix}
			& 			&\\
&\frac{e^{-i\frac{2\pi}{N}n j}}{\sqrt N} 	&\\
			&			&\\
\end{pmatrix}
\begin{pmatrix}
\ddots 		& 			&\\
&e^{-i\frac{K}{\hbar}\cos(\frac{2\pi}{N}j-\frac{\varphi}{N})}	&\\
			&			&\ddots\\
\end{pmatrix}
\begin{pmatrix}
			& 			&\\
&\frac{e^{i\frac{2\pi}{N}j m}}{\sqrt N}	&\\
			&			&\\
\end{pmatrix}\\
&\qquad\qquad\qquad\begin{pmatrix}
\ddots 		& 			&\\
	&e^{-im\frac{\varphi}{N}}	&\\
			&			&\ddots\\
\end{pmatrix}\\
\end{split}
\end{equation}

\section{Calculation of symmetric $B$ matrix}
\label{symmetricB.apx}

$D_{1}$ is a diagonal unitary matrix and $F$ is a unitary matrix.
The corresponding matrix elements are
$\left[D_{1}\right]_{n,n}=e^{i\frac{2\pi-\hbar}{2}n^2}$,
$F_{m,n}=\frac{1}{\sqrt N}e^{i\frac{2\pi}{N}mn}$,
where $\hbar=2\pi\frac{M}{N}$, and indices $m$ and $n$ take values $0,1,\cdots,N-1$.  We also
assume $k$ is an integer ranging from 1 to $N$, and $\tilde{k}$ is an integer ranging from
1 to $Q$, with $Q=(N-1)/2$.   From $B\equiv FD_1F^{\dag}$ and using the fact that $MN$ is an odd number,
we have
\begin{equation}
\begin{split}
B_{m,n}=&\frac{1}{N}\sum_{k=1}^{N}e^{i\frac{2\pi}{N}\left[\frac{M}{2}k^2+(\frac{N}{2}+m-n)k\right]}\\
=&\frac{1}{N}\sum_{k=1}^{N}(-1)^ke^{i\frac{2\pi}{N}\left[\frac{M}{2}k^2+(m-n)k\right]}\\
=&\frac{1}{N}\sum_{\tilde k=1}^{Q}e^{i\frac{2\pi}{N}\left[\frac{M}{2}(2\tilde k)^2+(m-n)(2\tilde k)\right]}\\
&\quad-\frac{1}{N}\sum_{\tilde k=1}^{Q}e^{i\frac{2\pi}{N}\left[\frac{M}{2}(N-2\tilde k)^2+(m-n)(N-2\tilde k)\right]}\\
&\quad-\frac{1}{N}e^{i\frac{2\pi}{N}\left[\frac{M}{2}N^2+(m-n)N\right]}\\
=&\frac{1}{N}+ \frac{1}{N}\sum_{\tilde k=1}^{Q}e^{i\frac{2\pi}{N}\left[2M\tilde k^2+2(m-n)\tilde k\right]}\\
&\quad-\frac{1}{N}\sum_{\tilde k=1}^{Q}e^{i\frac{2\pi}{N}\left[2M\tilde k^2+\frac{M}{2}N^2-2MN\tilde k+(m-n)N-2(m-n)\tilde k\right]}\\
=&\frac{1}{N}+\frac{1}{N}\sum_{\tilde k=1}^{Q}\left[e^{i\frac{2\pi}{N}\left(2M\tilde k^2+2(m-n)\tilde k\right)}+e^{i\frac{2\pi}{N}\left(2M\tilde k^2-2(m-n)\tilde k\right)}\right]\\
=&\frac{1}{N}+\frac{2}{N}\sum_{\tilde k=1}^{Q}e^{i4\pi\frac{M}{N}\tilde k^2}\cos{\left[4\pi\frac{\tilde k}{N}(m-n)\right]}.
\end{split}
\end{equation}
It is now seen that $B$ is a symmetric matrix, i.e., $B_{m,n}=B_{n,m}$.

\end{document}